\documentclass[journal,letterpaper,onecolumn,11pt]{IEEEtran}
\usepackage{verbatim,setspace,cite}
\usepackage{graphicx}
\usepackage{maxim}

\def\cA{{\cal A}}

\def\cF{{\cal F}}

\def\cM{{\cal M}}
\def\cP{{\cal P}}
\def\cS{{\cal S}}
\def\cU{{\cal U}}
\def\cW{{\cal W}}
\def\cX{{\cal X}}

\def\sS{{\sf S}}
\def\sV{{\sf V}}

\def\hcX{\widehat{\cal X}}
\def\hX{\widehat{X}}

\def\bd#1{\mathbf{#1}}
\def\wh#1{\widehat{#1}}
\def\td#1{\widetilde{#1}}

\begin{document}

\title{Joint Fixed-Rate Universal Lossy Coding\\
and Identification of Continuous-Alphabet Memoryless Sources}

\author{Maxim Raginsky,~\IEEEmembership{Member,~IEEE}%
\thanks{The material in this paper was presented in part at the IEEE International symposium on Information Theory, Seattle, July 9 -- July 14, 2006. This work
    was supported by the Beckman Institute Fellowship.}%
  \thanks{M.~Raginsky is with the Beckman Institute for Advanced
    Science and Technology, University of
Illinois, Urbana, IL 61801 USA (e-mail:~maxim@uiuc.edu).} }%

\maketitle

\thispagestyle{empty}
\pagestyle{plain}

\begin{abstract}
\noindent The problem of joint universal source coding and identification is considered in the setting of fixed-rate lossy coding of continuous-alphabet memoryless sources. For a wide class of bounded distortion measures, it is shown that any compactly parametrized family of $\R^d$-valued i.i.d. sources with absolutely continuous distributions satisfying appropriate smoothness and Vapnik--Chervonenkis learnability conditions, admits a joint scheme for universal lossy block coding and parameter estimation, such that when the block length $n$ tends to infinity, the overhead per-letter rate and the distortion redundancies converge to zero as $O(n^{-1}\log n)$ and $O(\sqrt{n^{-1}\log n})$, respectively. Moreover, the active source can be determined at the decoder up to a ball of radius $O(\sqrt{n^{-1} \log n})$ in variational distance, asymptotically almost surely. The system has finite memory length equal to the block length, and can be thought of as blockwise application of a time-invariant nonlinear filter with initial conditions determined from the previous block. Comparisons are presented with several existing schemes for universal vector quantization, which do not include parameter estimation explicitly, and an extension to unbounded distortion measures is outlined. Finally, finite mixture classes and exponential families are given as explicit examples of parametric sources admitting joint universal compression and modeling schemes of the kind studied here.\\ \\
\noindent {\bf Keywords:} Learning, minimum-distance density estimation, two-stage codes, universal vector quantization, Vapnik--Chervonenkis dimension.
\end{abstract}

\section{Introduction}
\label{sec:intro}

In a series of influential papers \cite{Ris84,Ris86,Ris96}, Rissanen
has elucidated and analyzed deep connections between universal lossless coding and statistical modeling. His approach hinges on the following two key insights:
\begin{enumerate}
\item A given parametric class of information
sources admits universal lossless codes if (a) the statistics of each
source in the class (or, equivalently, the parameters of the source) can be determined with arbitrary precision from a
sufficiently long data sequence and if (b) the parameter space can be partitioned
into a finite number of subsets, such that the sources whose
parameters lie in the same subset are ``equivalent'' in the sense of requiring ``similar'' optimal coding
schemes. This idea extends naturally to hierarchical model classes (e.g., when the dimension of the parameter vector is unknown), provided that the parametric family of sources governed by each model satisfies the above regularity conditions individually.
\item Given a sequence of symbols emitted by an information source from a hierarchical model class, an asymptotically correct model of the source is obtained by finding the best trade-off between the number of bits needed to describe it and the number of bits needed to losslessly encode the data assuming that the data are drawn from the maximum-likelihood distribution relative to this model. This is the basis of the so-called Minimum Description Length (MDL) principle for model selection and, more generally, statistical inference (see, e.g., the survey article of Barron, Rissanen and Yu \cite{BarRisYu98} or the recent book by Gr\"unwald \cite{Gru07}).
\end{enumerate}
There is, in fact, a natural symmetry between these two insights, owing to the well-known
one-to-one correspondence between (almost) optimal lossless codes and
probability distributions on the space of all input sequences
\cite{CovTho91}. For this reason, when considering universal lossless coding, we can use the term ``model" to refer either to the probability distribution of the source or to an optimal lossless code for the source, where we allow codes with ideal (noninteger) codeword lenghts. The main point of Rissanen's approach is precisely that the objectives of source coding and modeling can be accomplished jointly and in an asymptotically optimal manner.

Consider the case of a parametric class of sources where the parameter space has finite dimension $k$. Then the redundancy of the corresponding universal lossless code (i.e., the excess average codelength relative to the optimal code for the actual source at a given block length) is controlled essentially by the number of bits required to describe the source parameters to the decoder. In particular, the achievability theorem of Rissanen \cite[Theorem~1b]{Ris84} states that one can use a scheme of this kind to achieve the redundancy of about $(k/2)\log n/n$ bits per symbol, where $n$ is the block length. The universal lossless coder used by Rissanen in \cite{Ris84} operates as
follows: first, the input data sequence is used to compute the maximum-likelihood estimate of the parameters of the source, then the
estimate is quantized to a suitable resolution, and finally the data are encoded with the
corresponding optimum
lossless code. Structurally, this is an example
of a {\em two-stage code}, in which the binary description of the input data sequence
produced by the encoder consists of two parts: the first part
describes the (quantized) maximum-likelihood estimate of the source
parameters, while the second part describes the data using the code
matched to the estimated source.

In this paper, we investigate achievable redundancies in schemes for joint source coding and identification (modeling) in the setting of fixed-rate universal {\em lossy} block coding (vector quantization) of
continuous-alphabet memoryless sources. Once we pass from lossless codes to lossy ones, the term ``model" can refer either to a probabilistic description of the source or to a probability distribution over codebooks in the {\em reproduction} space.  In particular, whereas choosing a lossless code for an information source is equivalent to choosing a probabilistic model of the source, choosing a {\em lossy} code corresponds in a certain sense to sampling from a discrete probability distribution over sequences in the {\em reproduction} alphabet, and is thus related to the {\em source} distribution only indirectly. To place the present work in a wider context,  in Section~\ref{sec:summary} we briefly comment on the line of research concerned with relating lossy codes to codebook models, which can be thought of as a lossy variant of the MDL principle. However, there are situations in which one would like to compress the source and identify its statistics at the same time. For instance, in {\em indirect adaptive control} (see, e.g., Chapter~7 of Tao \cite{Tao03}) the parameters of the plant (the controlled system) are estimated on the basis of observation, and the controller is modified accordingly. Consider the discrete-time stochastic setting, in which the plant state sequence is a random process whose statistics are governed by a finite set of parameters. Suppose that the controller is geographically separated from the plant and connected to it via a noiseless digital channel whose capacity is $R$ bits per use. Then, given the time horizon $T$, the objective is to design an encoder and a decoder for the controller to obtain reliable estimates of both the plant parameters and the plant state sequence from the $2^{TR}$ possible outputs of the decoder.  In this paper, we are concerned with modeling the actual source directly, and not through a codebook distribution in the reproduction space. 

The objective of universal lossy coding (see, e.g.,
\cite{Ziv72,NeuGraDav75,LinLugZeg94,LinLugZeg95,ZhaWei96,ChoEffGra96})
is to construct lossy block source codes (vector quantizers) that perform well in incompletely or
inaccurately specified statistical environments. Roughly speaking, a
sequence of vector quantizers is universal for a given class of
information sources if it has asymptotically optimal performance,
in the sense of minimizing the average distortion under the rate
constraint, on any source in the class. Two-stage codes have
also proved quite useful in universal lossy coding
\cite{LinLugZeg94,LinLugZeg95,ChoEffGra96}. For instance, the two-stage universal quantizer introduced by Chou, Effros and Gray \cite{ChoEffGra96} is similar in
spirit to the adaptive lossless coder of Rice and Plaunt
\cite{RicPla70,RicPla71}, known as the ``Rice machine'': each input data sequence is encoded in parallel with a
number of codes, where each code is matched to one of the finitely many ``representative''
sources, and the code that performs the best on the given sequence
(in the case of lossy codes, compresses it with the smallest amount of
distortion) wins. Similar to the setting of Rissanen's achievability theorem, the  approach of \cite{ChoEffGra96} assumes a sufficiently smooth dependence
of optimum
coding schemes on the parameters of the source. However, the decision
rule used in selection of the second-stage code does not rely on
explicit modeling of the source statistics as the second-stage code is
chosen on the basis of local (pointwise), rather than average,
behavior of the data sequence with respect to a fixed collection of quantizers. This approach emphasizes the coding
objective at the expense of the modeling objective, thus falling short of exhibiting a relation between the latter and the former.

In the present work, we consider parametric
spaces $\{P_\theta\}$ of i.i.d. sources with values in $\R^d$, such
that the $P_\theta$'s are absolutely continuous and the parameter
$\theta$ belongs to a bounded subset of $\R^k$. We show in a
constructive manner that, for a wide class of bounded distortion functions and under certain regularity conditions, such
parametric families admit universal sequences of quantizers with distortion
redundancies\footnote{The distortion redundancy of a lossy block code
  relative to a source is the excess distortion of the code compared to the
  optimum code for that source.} converging to zero as $O(\sqrt{n^{-1}\log n})$ and with an
  overhead per-letter rate converging to zero as $O(n^{-1}\log
  n)$, as the block length $n \to \infty$. These convergence rates are, more or less, typical for universal coding schemes relying on explicit or implicit acquisition of the statistical model of the source (cf. the discussion in Section~\ref{sec:discuss} of this paper). For unbounded distortion functions satisfying a certain moment
  condition with respect to a fixed reference letter, the distortion redundancies are shown to converge to zero
  as $O(\sqrt[4]{n^{-1}\log n})$. The novel feature of our method, however, is
  that the decoder can use the two-stage binary description of the data not only to reconstruct the data with
  asymptotically optimal fidelity, but also to identify the active
  source up to a variational ball of radius $O(\sqrt{n^{-1}\log n})$ with
  probability approaching unity. In fact, the universality and the
  rate of convergence of the
  compression scheme are directly tied to the performance of the
  source identification procedure.

While our approach parallels Rissanen's method for proving his achievability theorem in \cite{Ris84}, there are two important
differences with regard to both his work on lossless codes and
subsequent work by others on universal lossy codes. The first difference is that the maximum-likelihood estimate, which fits
naturally into the lossless framework, is no longer appropriate in the lossy case. In order to relate coding
to modeling, we require that the probability distributions of the
sources under consideration behave smoothly as functions of the
parameter vectors; for compactly parametrized sources with absolutely continuous probability
distributions, this smoothness condition is stated as a local
Lipschitz property in terms of the $L_1$ distance between the
probability densities of the sources and the Euclidean distance in the parameter space. For bounded distortion measures, this implies that the expected performance of the corresponding
optimum coding schemes also exhibits smooth dependence on the
parameters. (By contrast, Chou, Effros and Gray
\cite{ChoEffGra96} impose the smoothness condition directly on the
optimum codes. This point will be elaborated upon in
Section~\ref{sec:twostage}.) Now, one can construct examples of sources with absolutely continuous
probability distributions for which the maximum-likelihood estimate behaves rather
poorly in terms of the $L_1$ distance between the true and the
estimated probability densities \cite{DevGyo01}. Instead, we propose the use of the
so-called {\em minimum-distance estimate}, introduced by Devroye and
Lugosi \cite{DevLug96,DevLug97} in the context of kernel density estimation. The introduction of the minimum-distance estimate allows
us to draw upon the powerful machinery of Vapnik--Chervonenkis
theory (see, e.g., \cite{DevLug01} and
Appendix~\ref{app:vc} in this paper) both for estimating the convergence rates
of density estimates and distortion redundancies, as well as for
characterizing the classes of sources that admit joint universal coding
and identification schemes. The merging of
Vapnik--Chervonenkis techniques with two-stage coding further underscores the forward relation between statistical learning/modeling and universal lossy coding.

The second difference is that, unlike previously proposed schemes,
our two-stage code has nonzero memory length. The use of memory is dictated
by the need to force the code selection procedure to be blockwise causal and robust to local variations in
the behavior of data sequences produced by ``similar'' sources. For a
given block length $n$, the stream
of input symbols is parsed into contiguous blocks of length $n$,
and each block is quantized with a quantizer matched to the source with the parameters estimated from the preceding block. In other words, the coding
process can be thought of as blockwise application of a nonlinear
time-invariant filter with initial conditions determined by the
preceding block. In the terminology of Neuhoff and Gilbert
\cite{NeuGil82}, this is an instance of a {\em block-stationary causal
  source code}.
 
The remainder of the paper is organized as follows. In
Section~\ref{sec:prelims}, we
state the basic notions of universal lossy coding specialized to block
codes with finite memory. Two-stage codes with memory are introduced in
Section~\ref{sec:twostage} and placed in the context of statistical
modeling and parameter estimation. The main result of this paper,
Theorem~\ref{thm:wu}, is also stated and proved in
Section~\ref{sec:twostage}. Next, in Section~\ref{sec:discuss}, we
present comparisons of our two-stage coding technique with several
existing techniques, as well as discuss some generalizations and
extensions. In Section~\ref{sec:examples} we show that two well-known
types of parametric sources --- namely, mixture classes and exponential
families --- satisfy, under mild regularity requirements, the conditions of our main theorem and thus admit joint universal quantization and identification schemes. Section~\ref{sec:summary} offers a quick summary of the paper, together with a list of potential topics for future research. Appendix~\ref{app:vc} contains a telegraphic
summary of notions and results from Vapnik--Chervonenkis
theory. Appendices \ref{app:drf}, \ref{app:mismatch} and \ref{app:barmismatch_proof} are devoted to proofs of certain technical results used throughout the paper.

\section{Preliminaries}
\label{sec:prelims}

Let $\{X_i\}^\infty_{i=-\infty}$ be a memoryless stationary source
with alphabet $\cX$ (the {\em source alphabet}), i.e., the $X_i$'s are independent
and identically distributed (i.i.d.) random variables with values in
$\cX$. Suppose that the common distribution of the $X_i$'s
belongs to a given indexed class $\{ P_\theta : \theta \in \Theta\}$ of probability measures on $\cX$ (with respect to an appropriate $\sigma$-field). The distributions on the
$n$-blocks $X^n = (X_1,\cdots,X_n)$ will be denoted by
$P^n_\theta$. The superscript $n$ will be dropped whenever it is clear
from the argument, such as in $P_\theta(x^n)$. Expectations with
respect to the corresponding process distributions will be denoted by
$\E_\theta [\cdot]$, e.g., $\E_\theta[X^n]$. In this paper, we assume
that $\cX$ is a Borel subset of $\R^d$, although this qualification is
not required in the rest of the present section.

Consider coding $\{X_i\}$ into another process $\{\hX_i\}$
with alphabet $\hcX$ (the {\em reproduction alphabet}) by means of a
finite-memory stationary block code. Given any $m,n,t \in \Z$ with $m,n \ge 1$, let $X^n_m(t)$ denote the segment
$$
(X_{tn - m+1},X_{tn - m+2},\cdots,X_{tn})
$$
of $\{X_i\}$. When $n=m$, we shall abbreviate this notation to
$X^n(t)$; when $m=1$, we shall write $X_n(t)$; finally, when $t=1$, we
shall write $X^n_m$, $X^n$, $X_m$. A code with block length $n$ and
memory length $m$ [or an $(n,m)$-block code, for short] is then
described as follows. Each reproduction $n$-block $\hX^n(t)$, $t \in
\Z$, is a function of the corresponding source $n$-block $X^n(t)$, as
well as of $X^n_m(t-1)$, the $m$ source symbols immediately preceding
$X^n(t)$, and this function is independent of $t$:
$$
\hX^n(t) = C^{n,m}(X^n(t),X^n_m(t-1)), \qquad \forall t \in \Z.
$$
When the code has zero memory, i.e., $m=0$, we shall denote it more
compactly by $C^n$. The performance of the code is measured in terms
of a {\em single-letter distortion} (or {\em fidelity criterion}),
i.e., a measurable map $\map{\rho}{\cX \times \hcX}{\R^+}$. The loss
incurred in reproducing a string $x^n \in \cX^n$ by $\wh{x}^n \in
\hcX^n$ is given by
$$
\rho(x^n,\wh{x}^n) = \sum^n_{i=1}\rho(x_i,\wh{x}_i).
$$
When the statistics of the source are described by $P_\theta$, the average per-letter
distortion of $C^{n,m}$ is defined as
\begin{eqnarray*}
D_\theta(C^{n,m}) &\deq& \limsup_{k \to \infty} \frac{1}{k}
\E_\theta[\rho(X^k,\hX^k)]\\
&=& \limsup_{k \to \infty} \frac{1}{k} \sum^k_{i=1}
\E_\theta[\rho(X_i,\hX_i)],
\end{eqnarray*}
where the $\hX_i$'s are determined from the rule $\hX^n(t) = C^{n,m}(X^n(t),X^n_m(t-1))$ for all $t \in
\Z$. Since the source is i.i.d., hence stationary, for each $\theta
\in \Theta$, both the reproduction process $\{\hX_i\}$ and the pair
process $\{(X_i,\hX_i)\}$ are $n$-stationary, i.e., the vector
processes $\{X^n(t)\}^\infty_{t=-\infty}$ and $\{ (X^n(t),\hX^n(t))
\}^\infty_{t=-\infty}$ are stationary \cite{NeuGil82}. This implies
\cite{GraNeuOmu75} that
$$
D_\theta(C^{n,m}) = \frac{1}{n}\E_\theta[\rho(X^n,\hX^n)] =
\frac{1}{n} \sum^n_{i=1} \E_\theta[\rho(X_i,\hX_i)],
$$
where $\hX^n = C^{n,m}(X^n,X^n_m(0))$.

More specifically, we shall consider fixed-rate lossy block codes
(also referred to as vector quantizers). A fixed-rate lossy
$(n,m)$-block code is a pair $(f,\phi)$ consisting of an encoder
$\map{f}{\cX^n \times \cX^m}{\cS}$ and a decoder
$\map{\phi}{\cS}{\hcX^n}$, where $\cS \subset \{0,1\}^*$ is a collection of fixed-length binary strings. The quantizer function
$\map{C^{n,m}}{\cX^n \times \cX^m}{\hcX^n}$ is the composite map
$\phi \circ f$; we shall often abuse notation, denoting by
$C^{n,m}$ also the pair $(f,\phi)$. The number $R(C^{n,m}) =
n^{-1} \log |\cS|$ is called the {\em rate} of $C^{n,m}$, in bits per
letter (unless specified otherwise, all logarithms in this paper will be taken to base 2). The set $\Gamma = \{\phi(s) : s \in \cS\}$ is the {\em
  reproduction codebook} of $C^{n,m}$.

The optimum performance achievable on the source $P_\theta$ by {\em
  any} finite-memory code with block length $n$ is given by the {\em
  $n$th-order operational distortion-rate function} (DRF)
$$
\wh{D}^{n,*}_\theta(R) \deq \inf_m \inf_{C^{n,m}} \big\{
D_\theta(C^{n,m}) : R(C^{n,m}) \le R \big\},
$$
where the infimum is over all finite-memory block codes with block
length $n$ and with rate at most $R$ bits per letter. If we restrict
the codes to have zero memory, then the corresponding $n$th-order
performance is given by
$$
\wh{D}^n_\theta(R) \deq \inf_{C^n} \big\{ D_\theta(C^n) :
R(C^n) \le R \big\}.
$$
Clearly, $\wh{D}^{n,*}_\theta(R) \le \wh{D}^n_\theta(R)$. However,
as far as optimal performance goes, allowing nonzero memory length
does not help, as the following elementary lemma shows:

\begin{lemma}\label{lm:nomem} $\wh{D}^{n,*}_\theta(R) =
  \wh{D}^n_\theta(R)$.
\end{lemma}

\begin{proof} It suffices to show that $\wh{D}^n_\theta(R) \le
  \wh{D}^{n,*}_\theta(R)$. Consider an arbitrary $(n,m)$-block code
  $C^{n,m} = \phi \circ f$, $\map{f}{\cX^n \times
    \cX^m}{\cS}$, $\map{\phi}{\cS}{\hcX^n}$. We claim that there
  exists a zero-memory code $C^n_* = \phi_* \circ f_*$,
  $\map{f_*}{\cX^n}{\cS}$, $\map{\phi_*}{\cS}{\hcX^n}$, such
  that
$$
\rho(x^n,C^n_*(x^n)) \le \rho(x^n,C^{n,m}(x^n,z^m))
$$
for all $x^n \in \cX^n, z^m \in \cX^m$. Indeed, define $f_*$ as
the minimum-distortion encoder
$$
x^n \mapsto \argmin_{s \in
  \cS}\rho(x^n,\phi(s))
$$
for the reproduction codebook of $C^{n,m}$,
and let $\phi_*(s) = \phi(s)$. Then it is easy to see that $R(C^n_*) \le R(C^{n,m})$ and $D_\theta(C^n_*)
\le D_\theta(C^{n,m})$ for all $\theta \in \Theta$, and the lemma is proved.
\end{proof}

Armed with this lemma, we can compare the performance of all
fixed-rate quantizers with block length $n$, with or without memory,
to the $n$th-order operational DRF $\wh{D}^n_\theta(R)$. If we allow
the block length to grow, then the best performance that can be
achieved by a fixed-rate quantizer with or without memory on the
source $P_\theta$ is given by the {\em operational distortion-rate
  function}
$$
\wh{D}_\theta(R) \deq \inf_n \wh{D}^n_\theta(R) = \lim_{n \to
  \infty} \wh{D}^n_\theta(R).
$$
Since an i.i.d. source is stationary and ergodic, the source coding
theorem and its converse \cite[Ch.~9]{Gal68} guarantee that the operational DRF
$\wh{D}_\theta(R)$ is equal to the {\em Shannon DRF} $D_\theta(R)$,
which in the i.i.d. case admits the following single-letter characterization:
$$
D_\theta(R) \deq \inf_Q \big\{ \E_{P_\theta Q}[\rho(X,\hX)] :
I_{P_\theta Q}(X,\hX) \le R \big\}.
$$
Here, the infimum is taken over all conditional probabilities (or test
channels) $Q$ from $\cX$ to $\hcX$, so that $P_\theta Q$ is the
corresponding joint probability on $\cX \times \hcX$, and $I$ is the
mutual information.

A universal lossy coding scheme at rate $R$ for the class $\{P_\theta
: \theta \in \Theta\}$ is a sequence of codes $\{C^{n,m}\}$, where
$n=1,2,\cdots$ and $m$ is either a constant or a function of $n$, such
that for each $\theta \in \Theta$, $R(C^{n,m})$ and
$D_\theta(C^{n,m})$ converge to $R$ and $D_\theta(R)$, respectively,
as $n \to \infty$. Depending on the mode of convergence with respect
to $\theta$, one gets different types of universal
codes. Specifically, let $\{C^{n,m}\}^\infty_{n=1}$ be a sequence of
lossy codes satisfying $R(C^{n,m}) \to R$ as $n \to \infty$. Then,
following \cite{NeuGraDav75}, we can distinguish between the following
three types of universality:

\begin{definition}[weighted universal]\label{def:wu} $\{C^{n,m}\}^\infty_{n=1}$ is {\em weighted
    universal} for $\{P_\theta : \theta \in \Theta\}$ with respect to
  a probability distribution $W$ on $\Theta$ (on an appropriate
  $\sigma$-field) if the {\em distortion redundancy}
$$
\delta_\theta(C^{n,m}) \deq D_\theta(C^{n,m}) - D_\theta(R)
$$
converges to zero in the mean, i.e.,
$$
\lim_{n \to \infty} \int_\Theta \delta_\theta(C^{n,m})dW(\theta) = 0.
$$
\end{definition}

\begin{definition}[weakly minimax universal]\label{def:wmu}
  $\{C^{n,m}\}^\infty_{n=1}$ is {\em weakly minimax universal} for
  $\{P_\theta : \theta \in \Theta \}$ if
$$
\lim_{n \to \infty} \delta_\theta(C^{n,m}) = 0
$$
for each $\theta \in \Theta$, i.e., $\delta_\theta(C^{n,m})$ converges
to zero pointwise in $\theta$.
\end{definition}

\begin{definition}[strongly minimax universal]\label{def:smu}
  $\{C^{n,m}\}^\infty_{n=1}$ is {\em strongly minimax universal} for
  $\{P_\theta : \theta \in \Theta\}$ if the convergence of
  $\delta_\theta(C^{n,m})$ to zero as $n \to \infty$ is uniform in $\theta$.
\end{definition}

The various relationships between the three types of universality have
been explored in detail, e.g., in \cite{NeuGraDav75}. From the
practical viewpoint, the differences between them are rather insubstantial. For instance, the existence of a weighted universal sequence of
codes for $\{P_\theta : \theta \in \Theta\}$ with respect to $W$
implies, for any $\epsilon > 0$ the existence of
a strongly minimax universal sequence for $\{P_\theta : \theta \in
\Theta_\epsilon\}$ for some $\Theta_\epsilon \subseteq \Theta$
satisfying $W(\Theta_\epsilon) \ge 1-\epsilon$. In this paper, we
shall concentrate exclusively on weakly minimax universal codes.

Once the existence of a universal sequence of codes is established in an
appropriate sense, we can proceed to determine the rate of
convergence. To facilitate this, we shall follow Chou, Effros and Gray
\cite{ChoEffGra96} and split the redundancy $\delta_\theta(C^{n,m})$
into two nonnegative terms:
\begin{equation}
\delta_\theta(C^{n,m}) = \big( D_\theta(C^{n,m}) - \wh{D}^n_\theta(R)
\big) + \big( \wh{D}^n_\theta(R) - D_\theta(R) \big).
\label{eq:redundancy}
\end{equation}
The first term, which we shall call the {\em $n$th-order redundancy}
and denote by $\delta^n_\theta(C^{n,m})$, quantifies the difference
between the performance of $C^{n,m}$ and the $n$th-order operational
DRF, while the second term tells us by how much the $n$th-order
operational DRF exceeds the Shannon DRF, with respect to the source
$P_\theta$. Note that $\delta_\theta(C^{n,m})$ converges to zero if
and only if $\delta^n_\theta(C^{n,m})$ does, because
$\wh{D}^n_\theta(R) \to D_\theta(R)$ as $n \to \infty$ by the source
coding theorem. Thus, in proving the existence of universal codes, we shall determine the rates at which the two terms on the right-hand side of (\ref{eq:redundancy}) converge to zero as $n \to \infty$.

\section{Two-stage joint universal coding and modeling}
\label{sec:twostage}

As discussed in the Introduction, two-stage codes are both practically
and conceptually appealing for analysis and design of universal
codes. A two-stage lossy block code (vector quantizer) with block
length $n$ is a code that
describes each source sequence $x^n$ in two stages: in the first stage, a
quantizer of block length $n$ is chosen as a function of $x^n$ from
some collection of available quantizers; this is followed by the
second stage, in which $x^n$ is encoded with the chosen code.

In precise terms, a two-stage fixed-rate lossy code is defined as follows
\cite{ChoEffGra96}. Let $\map{\td{f}}{\cX^n}{\td{\cS}}$ be a
mapping of $\cX^n$ into a collection $\td{\cS}$ of fixed-length
binary strings, and assume that to each $\td{s} \in \td{\cS}$ there
corresponds an $n$-block code $C^n_{\td{s}} = (f_{\td{s}},\phi_{\td{s}})$ at rate of $R$ bits per
letter. A two-stage code $C^n$ is defined by the encoder
$$
f(x^n) = \td{f}(x^n) f_{\td{f}(x^n)}(x^n)
$$
and the decoder
$$
\phi(\td{f}(x^n)f_{\td{f}(x^n)}(x^n)) =
\phi_{\td{f}(x^n)}(x^n).
$$
Here the juxtaposition of two binary strings stands for their concatenation. The map $\td{f}$ is called the first-stage
encoder. The rate of this code is $R + n^{-1} \log |\td{\cS}|$ bits per letter,
while the instantaneous distortion is
$$
\rho(x^n,C^n(x^n)) = \rho(x^n,C^n_{\td{f}(x^n)}(x^n)).
$$
Now consider using $C^n$ to code an i.i.d. process $\{X_i\}$ with
all the $X_i$'s distributed according to $P_\theta$ for some $\theta \in \Theta$. This will result in the
average per-letter distortion
\begin{eqnarray*}
D_\theta(C^n) &=& \frac{1}{n} \E_\theta[\rho(X^n,C^n(X^n))] \\
&=& \frac{1}{n} \E_\theta[\rho(X^n,C^n_{\td{f}(X^n)}(X^n))] \\
&=& \frac{1}{n} \int \rho(x^n,C^n_{\td{f}(x^n)}(x^n))
dP_\theta(x^n).
\end{eqnarray*}
Note that it is not possible to express $D_\theta(C^n)$ in terms of
expected distortion of any single code because the identity of the code
used to encode each $x^n \in  \cX^n$ itself varies with $x^n$.

Let us consider the following modification of two-stage
coding. As before, we wish to code an i.i.d. source $\{X_i\}$ with an $n$-block
lossy code, but this time we allow the code to have finite memory
$m$. Assume once again that we have an indexed collection $\{
C^n_{\td{s}} : \td{s} \in \td{\cS}\}$
of $n$-block codes, but this time the first-stage encoder is a map
$\map{\td{f}}{\cX^m}{\td{\cS}}$ from the space $\cX^m$ of $m$-blocks
over $\cX$ into $\td{\cS}$.  In order to encode the current $n$-block $X^n(t)$, $t \in \Z$, the
encoder first looks at $X^n_m(t-1)$, the $m$-block immediately
preceding $X^n(t)$, selects a code $C^n_{\td{s}}$ according to the
rule $\td{s} = \td{f}(X^n_m(t-1))$, and then codes $X^n(t)$ with
that code. In this way, we have a two-stage $(n,m)$-block code $C^{n,m}$
with the encoder
$$
f(x^n,z^m) = \td{f}(z^m)f_{\td{f}(z^m)}(x^n)
$$
and the decoder
$$
\phi(\td{f}(z^m)f_{\td{f}(z^m)}(x^n)) =
\phi_{\td{f}(z^m)}(x^n).
$$
The operation of this code can be pictured as a blockwise application
of a nonlinear time-invariant filter with the initial conditions
determined by a fixed finite amount of past data. Just as in the
memoryless case, the rate of $C^{n,m}$ is $R + n^{-1}\log|\td{\cS}|$ bits per
letter, but the instantaneous distortion is now given by
$$
\rho(x^n,C^{n,m}(x^n,z^m)) = \rho(x^n,C^n_{\td{f}(z^m)}(x^n)).
$$
When the common distribution of the $X_i$'s is $P_\theta$, the average
per-letter distortion is given by
\begin{eqnarray}
D_\theta(C^{n,m}) &=& \frac{1}{n} \int_{\cX^n \times \cX^m} \rho(x^n,C^n_{\td{f}(z^m)}(x^n)) dP_\theta(x^n,z^m) \nonumber \\
&=& \frac{1}{n} \E_\theta \left\{
  \E_\theta\left[\rho(X^n,C^n_{\td{f}(Z^m)}(X^n)) \Big| Z^m\right] \right\}
  \nonumber \\
&=& \E_\theta \left[ D_\theta\left(C^n_{\td{f}(X^m)}\right) \right].
\label{eq:2stdist_mem}
\end{eqnarray}
Observe that the use of memory allows us to decouple the choice of the code
from the actual encoding operation, which in turn leads to an
expression for the average distortion of $C^{n,m}$ that involves
iterated expectations.

\begin{figure}
\begin{center}
\includegraphics[width=0.6\columnwidth]{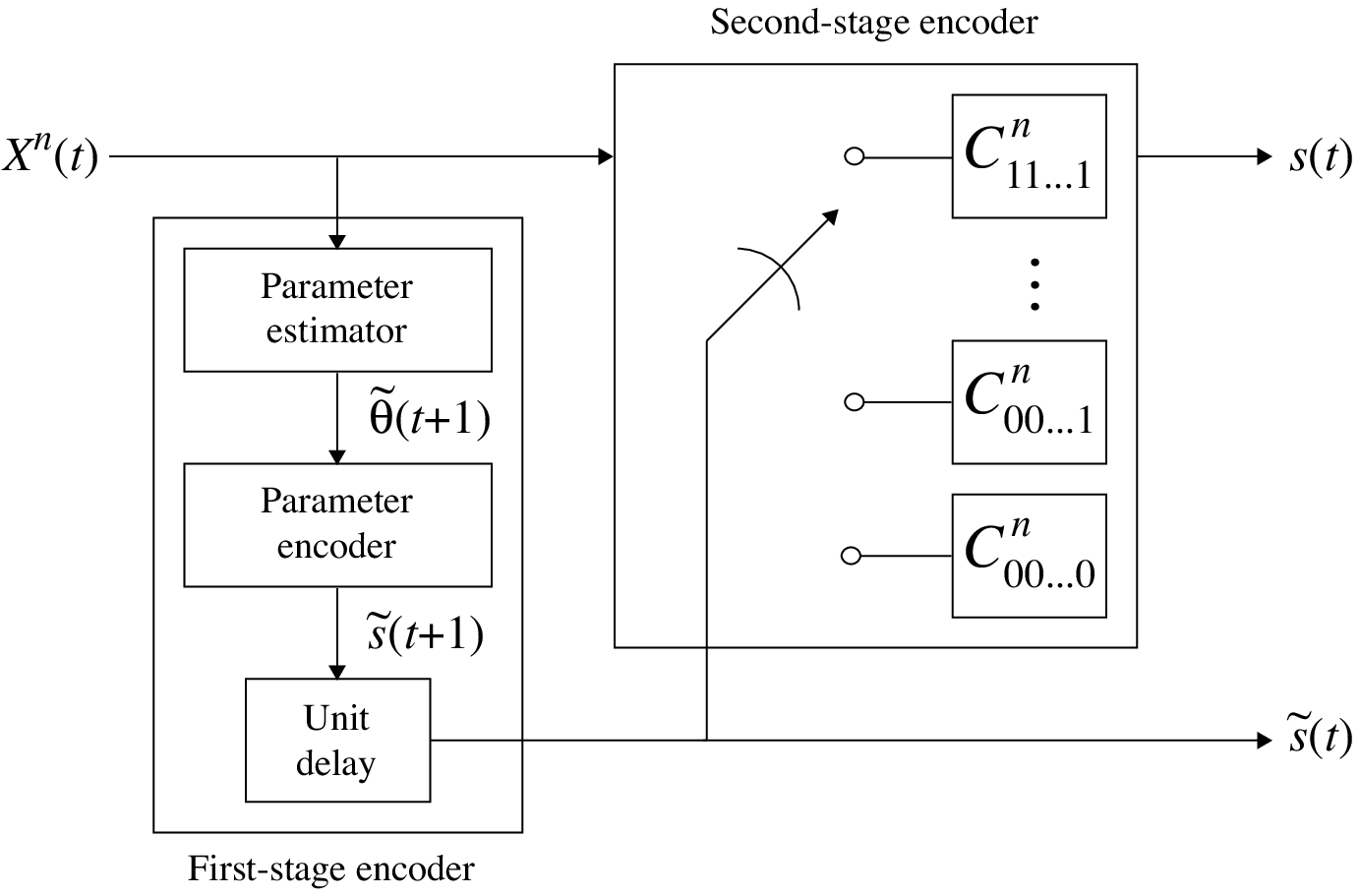} \\[12pt]
\includegraphics[width=0.6\columnwidth]{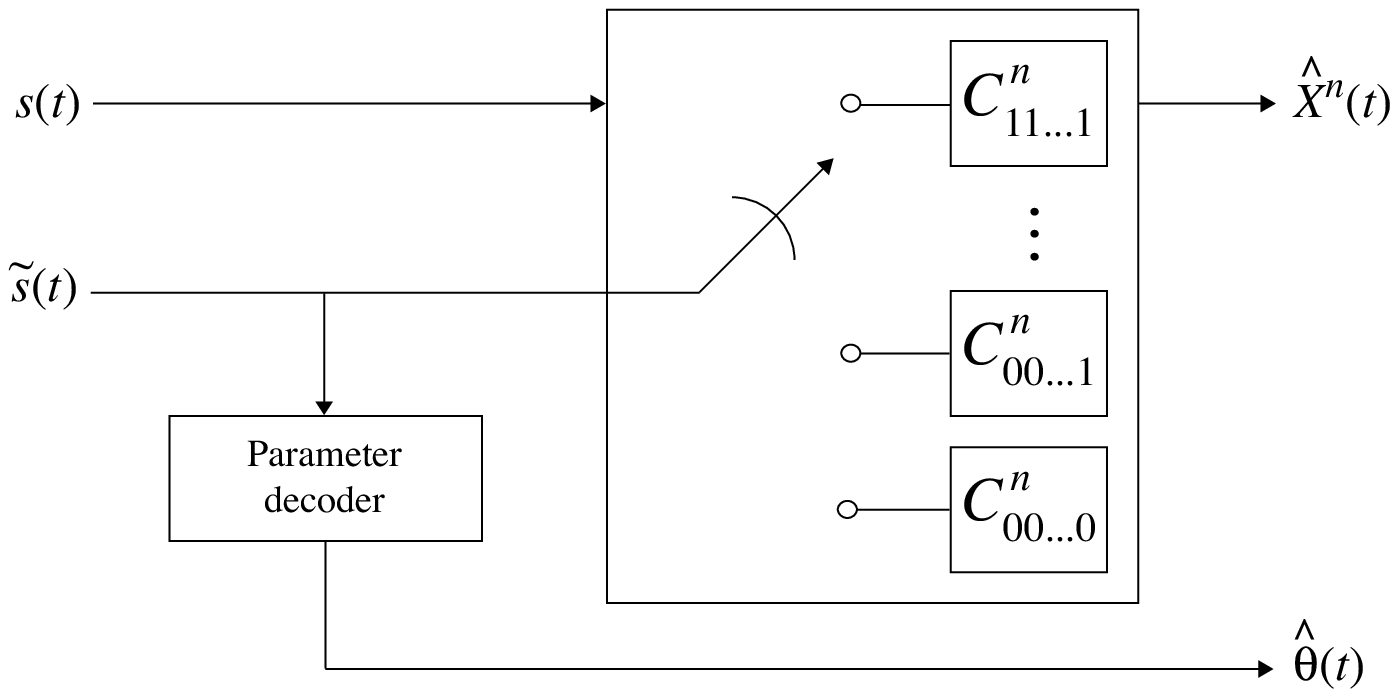}
\caption{\label{fig:twostage} Two-stage scheme for joint modeling and lossy coding (top: encoder; bottom: decoder).}
\end{center}\end{figure}

Intuitively, this scheme will yield a universal code if
\begin{equation}
\E_\theta \left[ D_\theta \left(C^n_{\td{f}(X^m)}\right) \right]
\approx \wh{D}^n_\theta(R)
\label{eq:goodcode}
\end{equation}
for each $\theta \in \Theta$. Keeping in mind that $\td{f}(X^m)$ is
allowed to take only a finite number $|\td{\cS}|$ of values, we see
that condition (\ref{eq:goodcode}) must be achieved through some combination of
parameter estimation and quantization. To this end, we impose
additional structure on the map $\td{f}$. Namely, we assume that it is
composed of a {\em parameter estimator}
$\map{\td{\theta}}{\cX^m}{\Theta}$ that uses the past data
$X^n_m(t-1)$ to estimate the parameter label $\theta \in \Theta$
of the source in effect, and a lossy {\em parameter encoder}
$\map{\td{g}}{\Theta}{\td{\cS}}$, whereby the estimate
$\td{\theta}$ is quantized to $\td{R}
\equiv \log |\td{\cS}|$ bits, with respect to a suitable distortion measure on $\Theta$. A binary description of the quantized version
$\wh{\theta}$ of $\td{\theta}(X^n_m(t-1))$ is then passed on to
the second-stage encoder which will quantize the current $n$-block
$X^n(t)$ with an $n$-block code matched to
$P_{\wh{\theta}}$. Provided that $P_\theta$ and $P_{\wh{\theta}}$
are ``close'' to each other in an appropriate sense, the resulting performance will be
almost as good as if the actual parameter $\theta$ were known all
along. As a bonus, the decoder will also receive a good $\td{R}$-bit binary
representation (model) of the source in effect. Therefore, we shall
also define a {\em parameter decoder} $\map{\td{\psi}}{\td{\cS}}{\Theta}$, so that
  $\wh{\theta} = \td{\psi}(\td{f}(X^n_m(t-1))$ can be taken as an estimate of the
  parameter $\theta \in \Theta$ of the active source. The structure of the encoder and the decoder in this two-stage scheme for joint modeling and lossy coding is displayed in Fig.~\ref{fig:twostage}.
  
These ideas are formalized in Theorem~\ref{thm:wu} below for i.i.d. vector sources $\{X_i\}$, $X_i \in \R^d$, where the common distribution of the $X_i$'s is a member of a given indexed class $\{P_\theta : \theta \in \Theta\}$ of absolutely continuous distributions, and the parameter space $\Theta$ is a bounded subset of $\R^k$. For simplicity we have set $m=n$, although other choices for the memory length are
also possible. Before we state and prove the theorem, let us fix some  useful results and notation. The following proposition generalizes Theorem 2 of Linder, Lugosi and Zeger \cite{LinLugZeg95} to i.i.d. vector sources and characterizes the rate at which the $n$th-order operational DRF converges to the Shannon DRF (the proof, which uses \Csiszar's generalized parametric representation of the DRF \cite{Csi74}, as well as a combination of standard random coding arguments and large-deviation estimates, is an almost verbatim adaptation of the proof of Linder {\em et al.} to vector sources, and is presented for completeness in Appendix~\ref{app:drf}):

\begin{proposition}\label{prop:drf} Let $\{X_i\}$ be an i.i.d. source with alphabet $\cX \subseteq \R^d$, where the common distribution of the $X_i$'s comes from an indexed class $\{P_\theta : \theta \in \Theta\}$. Let $\map{\rho}{\cX \times \hcX}{\R^+}$ be a distortion function satisfying the following two conditions:
\begin{enumerate}
\item $\Inf_{\wh{x} \in \hcX} \rho(x,\wh{x}) = 0$ for all $x \in \cX$.
\item $\Sup_{x \in \cX, \wh{x} \in \hcX}\rho(x,\wh{x}) = \rho_{\max} < \infty$.
\end{enumerate}
Then for every $\theta \in \Theta$ and every $R > 0$ such that $D_\theta(R) > 0$ there exists a constant $c_\theta(R)$ such that
$$
\wh{D}^n_\theta(R) - D_\theta(R) \le (c_\theta(R) + o(1)) \sqrt{\frac{\log n}{n}}.
$$
The function $c_\theta(R)$ is continuous in $R$, and the $o(1)$ term converges to zero uniformly in a sufficiently small neighborhood of $R$.
\end{proposition}

\begin{remark}The condition $D_\theta(R) > 0$ is essential to the proof and holds for all $R > 0$ whenever $P_\theta$ has a continuous component, which is assumed in the following.\end{remark}

\begin{remark}The constant $c_\theta(R)$ depends on the derivative of the DRF $D_\theta(R)$ at $R$ and on the maximum value $\rho_{\max}$ of the distortion function.\\[-10pt]\end{remark}
The distance between two i.i.d. sources will be measured by the
{\em variational distance} between their respective single-letter
distributions \cite[Ch.~5]{DevLug01}:
$$
d_V(P_\theta,P_\eta) \deq \sup_B |P_\theta(B) - P_\eta(B)|,\qquad
\forall \theta,\eta \in \Theta
$$
where the supremum is taken over all Borel subsets of $\R^d$. Also,
given a sequence of real-valued random variables $V_1,V_2,\cdots$ and a sequence
of nonnegative numbers $a_1,a_2,\cdots$, the notation $V_n = O(a_n)$
a.s. means that there exist a constant $c>0$ and a nonnegative random
variable $N \in \Z$ such that $V_n \le ca_n$ for all $n \ge N$. Finally,
both the statement and the proof of the theorem rely on certain
notions from Vapnik--Chervonenkis theory; for the reader's
convenience, Appendix~\ref{app:vc} contains a summary of the necessary
definitions and results.

\begin{theorem}\label{thm:wu} Let $\{X_i\}^\infty_{i=-\infty}$ be an i.i.d. source with alphabet $\cX \subseteq \R^d$, where the common distribution of the $X_i$'s is a member of a class $\{P_\theta : \theta \in \Theta\}$ of absolutely continuous distributions with the corresponding densities $p_\theta$. Assume the following conditions are satisfied:

\begin{enumerate}
\item $\Theta$ is a bounded subset of $\R^k$.

\item The map $\theta \mapsto P_\theta$ is uniformly locally Lipschitz: there exist constants $r > 0$ and $m > 0$ such that, for each
  $\theta \in \Theta$,
$$
d_V(P_\theta,P_\eta) \le m \| \theta - \eta
\|
$$
for all $\eta \in B_r(\theta)$, where $\| \cdot \|$ is the Euclidean norm on $\R^k$ and $B_r(\theta)$ is an open ball of radius $r$ centered at $\theta$.

\item The {\em Yatracos class} \cite{Yat85,DevLug96,DevLug97} associated with $\Theta$, defined as
$$
\cA_\Theta \deq \Big\{ A_{\theta,\eta} = \left\{x \in \cX: p_\theta(x) >
p_\eta(x) \right\} : \theta,\eta \in \Theta; \theta \neq \eta \Big\},
$$
is a Vapnik--Chervonenkis class, $\sV(\cA_\Theta) = V < \infty$.

\end{enumerate}
Let $\map{\rho}{\cX \times
  \hcX}{\R^+}$ be a single-letter distortion function of the form $\rho(x,\wh{x}) = [d(x,\wh{x})]^p$ for some $p > 0$, where $d(\cdot,\cdot)$ is a bounded metric on $\cX \cup \hcX$. Suppose that for each $n$ and each
$\theta \in \Theta$ there exists an $n$-block code $C^n_\theta =
(f_\theta,\phi_\theta)$ at rate of $R > 0$ bits per letter that
achieves the $n$th-order operational DRF for $P_\theta$:
$D_\theta(C^n_\theta) = \wh{D}^n_\theta(R)$. Then there exists an
$(n,n)$-block code $C^{n,n}$ with
\begin{equation}
R(C^{n,n}) = R + O\left(\frac{\log n}{n}\right),
\label{eq:rate}
\end{equation}
such that for every $\theta \in \Theta$
\begin{equation}
\delta_\theta(C^{n,n}) = O\left(\sqrt{\frac{\log n}{n}}\right).
\label{eq:dist}
\end{equation}
The resulting sequence of codes $\{C^{n,n}\}^\infty_{n=1}$ is therefore weakly
minimax universal for $\{P_\theta : \theta \in \Theta\}$ at rate
$R$. Furthermore, for each $n$ the first-stage encoder $\td{f}$ and  the corresponding parameter decoder
$\td{\psi}$ are such that
\begin{equation}
d_V(P_\theta,P_{\td{\psi}(\td{f}(X^n))}) = O\left( \sqrt{\frac{\log
    n}{n}} \right) \qquad {\rm a.s.,}
\label{eq:sourcedist}
\end{equation}
where the probability is with respect to $P_\theta$. The constants
implicit in the $O(\cdot)$ notation in (\ref{eq:rate}) and (\ref{eq:sourcedist}) are independent of $\theta$.
\end{theorem}

\begin{proof} The theorem will be proved by construction of a
  two-stage code, where the first-stage encoder
  $\map{\td{f}}{\cX^n}{\td{\cS}}$ is a cascade of the parameter
  estimator $\map{\td{\theta}}{\cX^n}{\Theta}$ and the lossy
  parameter encoder $\map{\td{g}}{\Theta}{\td{\cS}}$. Estimation of
  the parameter vector $\theta$ at the decoder will be facilitated by the
  corresponding decoder $\map{\td{\psi}}{\td{\cS}}{\Theta}$.

Our parameter estimator will be based on the so-called {\em minimum-distance
  density estimator} \cite[Sec.~5.5]{DevGyo01}, originally developed
by Devroye and Lugosi \cite{DevLug96,DevLug97} in the context of
kernel density estimation. It is constructed as follows. Let $Z^n =
(Z_1,\cdots,Z_n)$ be i.i.d. according to $P_\theta$ for some $\theta
\in \Theta$. Given any
$\eta \in \Theta$, let
$$
\Delta_\eta(Z^n) \deq \sup_{A \in \cA_\Theta} \left| P_\eta(A) -
P_{Z^n}(A) \right|,
$$
where $P_{Z^n}$ is the empirical distribution of $Z^n$,
$$
P_{Z^n}(B) \deq \frac{1}{n}\sum^n_{i=1}1_{\{Z_i \in B\}}
$$
for any Borel set $B$. Define $\td{\theta}(Z^n)$ as any $\theta^* \in \Theta$ satisfying
$$
\Delta_{\theta^*}(Z^n) < \inf_{\eta \in \Theta} \Delta_\eta(Z^n)
+ \frac{1}{n},
$$
where the extra $1/n$ term has been added to ensure that at least one
such $\theta^*$ exists. Then $P_{\td{\theta}(Z^n)}$ is called the
minimum-distance estimate of $P_\theta$. Through an abuse of
terminology, we shall also say that $\td{\theta}$ is the
minimum-distance estimate of $\theta$. The key property of the
minimum-distance estimate \cite[Thm.~5.13]{DevGyo01} is that
\begin{equation}
\int_\cX |p_{\td{\theta}(Z^n)}(x) - p_\theta(x)|dx \le 4
\Delta_\theta(Z^n) + \frac{3}{n}.
\label{eq:mde_error}
\end{equation}
Since the variational distance between any two
absolutely continuous distributions $P,Q$ on $\R^d$ is equal to one half of the
$L_1$ distance between their respective densities $p,q$ \cite[Thm.~5.1]{DevLug01}, i.e., 
$$
d_V(P,Q) = \frac{1}{2}\int_{\R^d} |p(x) - q(x)| dx,
$$
we can rewrite (\ref{eq:mde_error}) as
\begin{equation}
d_V(P_\theta,P_{\td{\theta}(Z^n)}) \le 2 \Delta_\theta(Z^n) + \frac{3}{2n}.
\label{eq:mindist}
\end{equation}
Since $\cA_\Theta$ is a Vapnik--Chervonenkis class,
Lemma~\ref{lm:vcclass_bound} in the Appendices asserts that
\begin{equation}
\E_\theta[\Delta_\theta(Z^n)] \le c_1\sqrt{\frac{\log n}{n}},
\label{eq:vcdelta}
\end{equation}
where $c_1$ is a constant that depends only on the VC dimension of
$\cA_\Theta$. Taking expectations of both sides of (\ref{eq:mindist})
and applying (\ref{eq:vcdelta}), we get
\begin{equation}
\E_\theta \left[d_V(P_\theta,P_{\td{\theta}(Z^n)})\right] \le 2c_1\sqrt{\frac{\log
    n}{n}} + \frac{3}{2n}.
\label{eq:mindist_2}
\end{equation}

Next, we construct the lossy encoder $\td{g}$. Since $\Theta$ is bounded, it is contained in some hypercube $M$ of side
$J$, where $J$ is some positive integer. Let $\cM^{(n)}
= \{M^{(n)}_1,M^{(n)}_2,\cdots,M^{(n)}_K\}$ be a partitioning of $M$
into nonoverlapping hypercubes of side $1/\lceil n^{1/2} \rceil$, so
that $K \le (Jn^{1/2})^k$. Represent each $M^{(n)}_j$ that intersects
$\Theta$ by a unique fixed-length binary string $\td{s}_j$, and let
$\td{\cS} = \{\td{s}_j\}$.  Then if a given $\theta \in \Theta$ is
contained in $M^{(n)}_j$, map it to $\td{s}_j$, $\td{g}(\theta) =
\td{s}_j$; this choice can be described by a string of no more than $k(\log n^{1/2} + \log J)$
bits. Finally, for each $M^{(n)}_j$ that intersects
$\Theta$, choose a representative $\wh{\theta}_j \in M^{(n)}_j \cap
\Theta$ and define the corresponding $n$-block code $C^n_{\td{s}_j}$ to be
$C^n_{\wh{\theta}_j}$. Thus, we can associate to $\td{g}$ the decoder
$\map{\td{\psi}}{\td{\cS}}{\Theta}$ via $\td{\psi}(\td{s}_j) = \wh{\theta}_j$.

Now let us describe and analyze the operation of the resulting two-stage
$(n,n)$-block code $C^{n,n}$. In
order to keep the notation simple, we shall suppress the discrete time
variable $t$ and denote the current block $X^n(t)$ by $X^n$, while the
preceding block $X^n(t-1)$ will be denoted by $Z^n$. The first-stage
encoder $\td{f}$ computes the minimum-distance
estimate $\td{\theta} = \td{\theta}(Z^n)$ and communicates its lossy binary description
$\td{s} = \td{f}(Z^n) \equiv \td{g}(\td{\theta}(Z^n))$ to the second-stage
encoder. The second-stage encoder then encodes $X^n$ with the
$n$-block code
$C^n_{\td{s}} \equiv C^n_{\wh{\theta}}$, where $\wh{\theta} =
\td{\psi}(\td{s})$ is the quantized version of the minimum-distance
estimate $\td{\theta}$. The string transmitted to the decoder thus
consists of two parts: the header $\td{s}$, which specifies the
second-stage code $C^n_{\td{s}}$, and the body $s = f_{\td{s}}(X^n)$,
which is the encoding of $X^n$ under $C^n_{\td{s}}$. The decoder
computes the reproduction $\hX^n = \phi_{\td{s}}(s)$. Note, however,
that the header $\td{s}$ not only instructs the
decoder how to decode the body $s$, but also contains a binary description
of the quantized minimum-distance estimate of the active source,
which can be recovered by means of the rule $\wh{\theta} =
\td{\psi}(\td{s})$.

In order to keep the notation simple, assume for now that $p=1$, i.e., the distortion function $\rho$ is a metric on $\cX \cup \hcX$ with the bound $\rho_{\max}$; the case of general $p$ is similar. The rate of $C^{n,n}$ is clearly no more than
$$
R + \frac{k(\log n^{1/2} + \log J)}{n}
$$
bits per letter, which proves (\ref{eq:rate}). The average-per letter distortion of $C^{n,n}$ on the source
$P_\theta$ is, in accordance with (\ref{eq:2stdist_mem}),  given by
$$
D_\theta(C^{n,n}) = \E_\theta \left[ D_\theta\left (C^n_{\td{f}(Z^n)}
  \right) \right],
$$
where $C^n_{\td{f}(Z^n)} = C^n_{\wh{\theta}}$ with $\wh{\theta} =
\td{\psi}(\td{f}(Z^n))$. Without loss of generality, we may assume that each $C^n_\theta$ is a nearest-neighbor quantizer, i.e.,
$$
\rho(x^n,C^n_\theta(x^n)) = \min_{\wh{x}^n \in \Gamma_\theta} \rho(x^n,\wh{x}^n)
$$
for all $x^n \in \cX^n$, where $\Gamma_\theta$ is the codebook of $C^n_\theta$. Then we have the following chain of estimates:
\begin{eqnarray*}
D_\theta(C^n_{\wh{\theta}}) & \stackrel{\rm(a)}{\le} &
D_{\wh{\theta}}(C^n_{\wh{\theta}}) + 2\rho_{\max}
d_V(P_\theta,P_{\wh{\theta}}) \nonumber \\
& \stackrel{\rm(b)}{=} & \wh{D}^n_{\wh{\theta}}(R) + 2\rho_{\max}
d_V(P_\theta,P_{\wh{\theta}}) \nonumber \\
& \stackrel{\rm(c)}{\le} & \wh{D}^n_\theta(R) + 4\rho_{\max}
d_V(P_\theta,P_{\wh{\theta}}) \nonumber \\
& \stackrel{\rm(d)}{\le} & \wh{D}^n_\theta(R) + 4\rho_{\max} \left[
  d_V(P_\theta,P_{\td{\theta}}) +
  d_V(P_{\td{\theta}},P_{\wh{\theta}}) \right],
\end{eqnarray*}
where (a) and (c) follow from a basic quantizer mismatch estimate
(Lemma~\ref{lm:qmismatch} in the Appendices), (b) follows from the assumed $n$th-order optimality of
$C^n_{\wh{\theta}}$ for $P_{\wh{\theta}}$, while (d) is a routine
application of the triangle inequality. Taking
expectations, we get
\begin{equation}
D_\theta(C^{n,n}) \le \wh{D}^n_\theta(R) + 4\rho_{\max}
\left\{\E_\theta [d_V(P_\theta,P_{\td{\theta}})] + \E_\theta
  [d_V(P_{\td{\theta}},P_{\wh{\theta}})]\right\}.
\label{eq:distbound}
\end{equation}
We now estimate separately each term in the curly brackets in (\ref{eq:distbound}). The first term can be bounded using the fact that
$\td{\theta} = \td{\theta}(Z^n)$ is a minimum-distance estimate of
$\theta$, so by (\ref{eq:mindist_2}) we have
\begin{equation}
\E_\theta [d_V(P_\theta,P_{\td{\theta}})] \le 2c_1 \sqrt{\frac{\log n}{n}}
+ \frac{3}{2n}.
\label{eq:1st_term}
\end{equation}
The second term involves $\td{\theta}$ and its quantized version
$\wh{\theta}$, which satisfy $\| \td{\theta} - \wh{\theta} \| \le
\sqrt{k/n}$, by construction of the parameter space quantizer
$(\td{g},\td{\psi})$. By the uniform local Lipschitz property
of the map $\theta \mapsto P_\theta$, there exist constants $r > 0$ and $m > 0$, such that
$$
d_V(P_\eta,P_{\wh{\theta}}) \le m \|
\eta - \wh{\theta} \|
$$
for all $\eta \in B_r(\wh{\theta})$. If $\td{\theta} \in
B_r({\wh{\theta}})$, this implies that
$d_V(P_{\td{\theta}},P_{\wh{\theta}}) \le m \sqrt{k/n}$. Suppose, on the other hand, that $\td{\theta} \not\in
B_r({\wh{\theta}})$. By assumption, $\| \td{\theta} - \wh{\theta} \| \ge
r$. Therefore, since $d_V(\cdot,\cdot)$ is bounded from above
by unity, we can write
$$
d_V(P_{\td{\theta}},P_{\wh{\theta}}) \le \frac{1}{r} \| \td{\theta} - \wh{\theta} \| \le \frac{1}{r}
\sqrt{\frac{k}{n}}.
$$
Let $b \deq \max(m,1/r)$. Then
the above argument implies that
\begin{equation}
d_V(P_{\td{\theta}},P_{\wh{\theta}})
\le b \sqrt{\frac{k}{n}}
\label{eq:2nd_term_noexp}
\end{equation}
and consequently
\begin{equation}
\E_\theta [d_V(P_{\td{\theta}},P_{\wh{\theta}})] \le b\sqrt{\frac{k}{n}}.
\label{eq:2nd_term}
\end{equation}
Substituting the bounds (\ref{eq:1st_term}) and (\ref{eq:2nd_term})
into (\ref{eq:distbound}) yields
$$
D_\theta(C^{n,n}) \le \wh{D}^n_\theta(R) + \rho_{\max} \left(
  8c_1\sqrt{\frac{\log n}{n}} + \frac{6}{n} + 4b \sqrt{\frac{k}{n}}
  \right),
$$
whence it follows that the $n$th-order redundancy $\delta^n_\theta(C^n) = O(\sqrt{n^{-1}\log n})$ for every $\theta \in \Theta$. Then the decomposition
$$
\delta_\theta(C^{n,n}) = \delta^n_\theta(C^{n,n}) + \wh{D}^n_\theta(R) - D_\theta(R)
$$
and Proposition~\ref{prop:drf} imply that (\ref{eq:dist}) holds for every $\theta \in \Theta$. The case of $p \neq 1$ is similar.

To prove (\ref{eq:sourcedist}), fix an $\epsilon > 0$ and note that by (\ref{eq:mindist}), (\ref{eq:2nd_term_noexp}) and the triangle inequality, $d_V(P_\theta,P_{\wh{\theta}(Z^n)}) >
\epsilon$ implies that
$$
2\Delta_\theta(Z^n) + \frac{3}{2n} + b\sqrt{\frac{k}{n}}  > \epsilon.
$$
Hence,
\begin{eqnarray*}
\Pr \left\{ d_V(P_\theta,P_{\wh{\theta}(Z^n)}) > \epsilon \right\} & \le & \Pr \left\{ \Delta_\theta(Z^n) > \frac{1}{2}\left(\epsilon -
\frac{3}{2n} - b\sqrt{\frac{k}{n}} \right) \right\} \\
&\le & \Pr \left\{ \Delta_\theta(Z^n) > \frac{1}{2}\left(\epsilon - c_2
\sqrt{\frac{1}{n}} \right) \right\},
\end{eqnarray*}
where $c_2 = 3/2 + b\sqrt{k}$. Therefore, by Lemma~\ref{lm:vcclass_bound},
\begin{equation}
\Pr \left\{ d_V(P_\theta,P_{\wh{\theta}(Z^n))}) > \epsilon \right\} \le
8n^V e^{-n(\epsilon - c_2 \sqrt{1/n})^2/128}.
\label{eq:lgdev}
\end{equation}
If for each $n$ we choose $\epsilon_n > \sqrt{128 V \ln n /n} + c_2 \sqrt{1/n}$, then the
right-hand side of (\ref{eq:lgdev}) will be summable in $n$, hence
$d_V(P_\theta,P_{\wh{\theta}(Z^n)}) = O(\sqrt{n^{-1}\log n})$ a.s. by the
Borel--Cantelli lemma.
\end{proof}

\begin{remark} Our proof combines the techniques of Rissanen \cite{Ris84}, in that the
  second-stage code is selected through explicit estimation
  of the source parameters, and of Chou, Effros and Gray \cite{ChoEffGra96}, in that the
  parameter space is quantized and each $\theta$ is identified with its optimal code $C^n_\theta$. The novel element here is
  the use of minimum-distance estimation instead of maximum-likelihood
  estimation, which is responsible for the appearance of the VC dimension.\end{remark}
  
\begin{remark} The boundedness of the distortion measure has been assumed mostly in order to ensure that the main idea behind the proof is not obscured by technical details. In Section~\ref{ssec:ubound} we present an extension to distortion measures that satisfy a moment condition with respect to a fixed reference letter in the reproduction alphabet. In that case, the parameter estimation fidelity and the per-letter overhead rate still converge to zero as $O(\sqrt{n^{-1}\log n})$ and $O(n^{-1}\log n)$, respectively, but the distortion redundancy converges more slowly as $O(\sqrt[4]{n^{-1}\log n})$. \end{remark}

\begin{remark} Essentially the same convergence rates, up to multiplicative and/or additive constants, can be obtained if the memory length is taken to be some fraction of the block length $n$: $m = \alpha n$ for some $\alpha \in (0,1)$. \end{remark}

\begin{remark} Let us compare the local Lipschitz condition of
  Theorem~\ref{thm:wu} to the corresponding smoothness conditions of
  Rissanen \cite{Ris84} for lossless codes and of Chou {\em et al.}
  \cite{ChoEffGra96} for quantizers. In the lossless case, $\cX$ is finite or countably infinite, and the smoothness condition is for the relative entropies $D(P_\eta
  \| P_\theta) \deq \sum_{x \in \cX} p_\eta(x) \log
  (p_\eta(x)/p_\theta(x))$, where $p_\theta$ and $p_\eta$ are the
  corresponding probability mass functions, to be locally quadratic in
  $\theta$: $D(P_\eta \| P_\theta) \le
  m_\theta \| \theta - \eta \|^2$ for some constant $m_\theta$ and for all $\eta$ in some open
  neighborhood of $\theta$. Pinsker's inequality
  $D(P_\eta \| P_\theta) \ge d^2_V(P_\eta,P_\theta)/2\ln 2$
  \cite[p.~58]{CsiKor81} then implies the local Lipschitz property for
  $d_V(P_\theta,P_\eta)$, although the magnitude of the Lipschitz constant is not uniform in $\theta$. Now, $D(P_\eta \| P_\theta)$ is
  also the
  redundancy of the optimum lossless code for $P_\theta$ relative to
  $P_\eta$. Thus, Rissanen's smoothness condition can be interpreted either in the context of source models or in
  the context of coding schemes and their redundancies. The latter interpretation has been extended
  to quantizers in \cite{ChoEffGra96}, where it was required that the redundancies $\delta^n_\eta(C^n_\theta)$ be locally quadratic in $\theta$. However, because here we are interested in
  joint modeling and coding, we impose a
  smoothness condition on the source distributions, rather than on the
  codes. The variational distance is more appropriate here than
  the relative entropy because, for bounded distortion functions, it
  is a natural measure of redundancy for lossy codes
  \cite{NeuGraDav75}.\end{remark}

\begin{remark} The Vapnik--Chervonenkis dimension of a given class of
  measurable subsets of $\R^k$ (provided it is finite) is, in a sense,
  a logarithmic measure of the combinatorial ``richness'' of the class for the
  purposes of learning from empirical data. For many parametric
  families of probability densities, the VC dimension of the
  corresponding Yatracos class is polynomial in $k$, the dimension of
  the parameter space (see \cite{DevGyo01} for detailed examples).\end{remark}

\begin{remark} Instead of the Vapnik--Chervonenkis condition, we could
  have required that the class of sources $\{P_\theta : \theta \in
  \Theta\}$ be {\em totally bounded} with respect to the variational distance. (Totally bounded classes, with respect to either the variational distance or its generalizations, such as the $\bar{\rho}$-distance \cite{GraNeuShi75}, have, in fact, been extensively used in the theory of universal lossy codes \cite{NeuGraDav75}.) This was precisely the assumption made in the
  paper of Yatracos \cite{Yat85} on density estimation, which in turn inspired the work of
  Devroye and Lugosi \cite{DevLug96,DevLug97}. The main result of Yatracos is that, if the class $\{P_\theta : \theta \in \Theta\}$ is totally bounded under the variational distance, then for any $\epsilon > 0$ there exists an estimator $\theta^* = \theta^*(X^n)$, where $X^n$ is an i.i.d. sample from one of the $P_\theta$'s, such that 
  $$
  \E_\theta[d_V(P_\theta,P_{\theta^*(X^n)})] \le 3\epsilon + \sqrt{\frac{32 H_\epsilon + 8}{n}},
  $$
  where $H_\epsilon$ is the metric entropy, or Kolmogorov $\epsilon$-entropy \cite{KolTih61}, of $\{P_\theta : \theta \in \Theta\}$, i.e., the logarithm of the cardinality of the minimal $\epsilon$-net for $\{P_\theta\}$ under $d_V(\cdot,\cdot)$. Thus, if we choose $\epsilon = \epsilon_n$ such that $\sqrt{H_{\epsilon_n}/n} \to 0$ as $n \to \infty$, then $\theta^*(X^n)$ is a consistent estimator of $\theta$. However, totally bounded classes have certain drawbacks. For example, depending on the structure and the complexity of the class, the Kolmogorov $\epsilon$-entropy may vary rather drastically from a polynomial in $\log (1/\epsilon)$ for ``small" parametric families (e.g., finite mixture families) to a polynomial in $1/\epsilon$ for nonparametric families (e.g., monotone densities on the hypercube or smoothness classes such as Sobolev spaces). One can even construct extreme examples of nonparametric families with $H_\epsilon$ {\em exponential} in $1/\epsilon$. (For details, the reader is invited to consult Ch.~7 of \cite{DevLug01}.) Thus, in sharp contrast to VC classes for which we can obtain $O(\sqrt{n^{-1} \log n})$ convergence rates both for parameter estimates and for distortion redundancies, the performance of joint universal coding and modeling schemes for totally bounded classes of sources will depend rather strongly on the metric properties of the class. Additionally, although in the totally bounded case there is no need for quantizing the parameter space, one has to construct an $\epsilon$-net for each given class, which is often an intractable problem.
\end{remark}

\section{Comparisons and extensions}
\label{sec:discuss}

\subsection{Comparison with nearest-neighbor and omniscient
  first-stage encoders}
\label{ssec:omnisci}

The two-stage universal quantizer of Chou, Effros and
Gray \cite{ChoEffGra96} has zero memory and works as
follows. Given a collection $\{C^n_{\td{s}}\}$ of $n$-block codes,
the first-stage encoder is given by the ``nearest-neighbor'' map
$$
\td{f}_*(x^n) \deq \argmin_{\td{s} \in \td{\cS}}
\rho(x^n,C^n_{\td{s}}(x^n)),
$$
where the term ``nearest-neighbor'' is used in the sense that the code $C^n_{\td{f}_*(x^n)}$ encodes $x^n$ with the smallest
instantaneous distortion among all $C^n_{\td{s}}$'s. Accordingly, the average
per-letter distortion of the resulting two-stage code $C^n_*$ on the
source $P_\theta$ is given by
$$
D_\theta(C^n_*) = \frac{1}{n}\int \min_{\td{s} \in \td{\cS}}
\rho(x^n,C^n_{\td{s}}(x^n)) dP_\theta(x^n).
$$
Although such a code is easily implemented in
practice, its theoretical analysis is quite complicated. However, the
performance of $C^n_*$ can be upper-bounded if the nearest-neighbor
first-stage encoder is replaced by the so-called {\em omniscient}
first-stage encoder, which has direct access to the source parameter
$\theta \in \Theta$, rather than to $x^n$. This latter encoder is obviously not achievable in
practice, but is easily seen to do no better than the nearest-neighbor
one.

This approach can be straightforwardly adapted to the
setting of our Theorem~\ref{thm:wu}, except that we no longer require
Condition 3). In that case, it is apparent that the sequence $\{C^n_*\}$
of the two-stage $n$-block (zero-memory) codes with nearest-neighbor
(or omniscient) first-stage encoders is such that
\begin{equation}
R(C^n_*) = R + O\left(\frac{\log n}{n}\right)
\label{eq:nnrate}
\end{equation}
and
\begin{equation}
\delta_\theta(C^n_*) = O\left(\sqrt{\frac{\log n}{n}}\right).
\label{eq:nndist}
\end{equation}
Comparing (\ref{eq:nnrate}) and (\ref{eq:nndist}) with (\ref{eq:rate}) and (\ref{eq:dist}), we immediately see that the
use of memory and direct parameter estimation has no effect on rate or on distortion. However, our scheme uses the $O(\log n)$ overhead bits in
a more efficient manner --- indeed, the bits produced by the
nearest-neighbor first-stage encoder merely tell the second-stage
encoder and the decoder which quantizer to use, but there is, in general,
no guarantee that the nearest-neighbor code for a given $x^n \in \cX^n$ will
be matched to the actual source in an average sense. By contrast, the
first-stage description under our scheme, while requiring essentially
the same number of extra bits, can be used to identify the acive
source up to a variational ball of radius $O(\sqrt{n^{-1}\log n})$, with probability arbitrarily close to one.

\subsection{Comparison with schemes based on codebook transmission}
\label{ssec:cbxmit}

Another two-stage scheme, due to Linder, Lugosi and Zeger
\cite{LinLugZeg94,LinLugZeg95}, yields weakly minimax universal codes
for all real i.i.d. sources with bounded support, with respect to the
squared-error distortion. The main feature of their approach is that, instead of
constraining the first-stage encoder to choose from a collection of
preselected codes, they encode each $n$-block $x^n \in \cX^n$ by designing, in real time, an optimal quantizer for the empirical distribution
$P_{x^n}$, whose codevectors are then quantized to some carefully chosen
resolution. Then, in the second stage, $x^n$ is quantized with this ``quantized quantizer,'' and a binary description of the quantized codevectors is transmitted together
with the second-stage description of $x^n$. The overhead needed to transmit the quantized
codewords is $O(n^{-1}\log n)$ bits per letter, while the distortion
redundancy converges to zero at a rate $O(\sqrt{n^{-1}\log n})$.

In order to draw a comparison with the results presented here, let
$\{ P_\theta : \theta \in \Theta\}$ be a class of real i.i.d. sources
satisfying Conditions 1)--3) of Theorem~\ref{thm:wu}, and with support
contained in some closed interval $[-B,B]$, i.e., $P_\theta\{|X|
\le B\} = 1$ for all $\theta \in \Theta$. Let also $\hcX = \R$, and consider the squared-error
distortion $\rho(x,\wh{x}) = |x - \wh{x}|^2$. Without loss of
generality, we may assume that the optimal $n$-block quantizers
$C^n_\theta$ have nearest-neighbor encoders, which in turn allows us
to limit our consideration only to those quantizers whose codevectors
have all their components  in $[-B,B]$. Then $\rho$ is bounded with
$\rho_{\max} = 4B^2$, and Theorem~\ref{thm:wu} guarantees the
existence of a weakly minimax universal sequence $\{C^{n,n}\}$ of
$(n,n)$-block codes satisfying (\ref{eq:rate}) and
(\ref{eq:dist}). Comparing this with the results of Linder {\em et al.} quoted in the preceding paragraph, we see that, as far as the
rate and the distortion redundancy go, our scheme performs as well as
that of \cite{LinLugZeg94,LinLugZeg95}, but, again, in our case the extra
$O(\log n)$ bits have been utilized more efficiently, enabling the
decoder to identify the active source with good precision. However,
the big difference between our code and that of Linder {\em et al.} is that the class of sources considered by them is {\em fully
  nonparametric}, whereas our development requires that the sources belong to a
compactly parametrized family.

\subsection{Extension to curved parametric families}
\label{ssec:curved}

We can also consider parameter spaces that are more general than bounded
subsets of $\R^k$. For instance, in information geometry
\cite{AmaNag00} one often
encounters {\em curved parametric families}, i.e., families $\{P_\theta : \theta \in \Theta\}$ of probability
distributions where the parameter space $\Theta$ is a smooth compact manifold. Roughly speaking,
an abstract set $\Theta$ is a smooth compact manifold of dimension
$k$ if it admits a covering by finitely many sets $G_l \subset
\Theta$, such that for each $l$ there exists a one-to-one
map $\xi_l$ of $G_l$ onto a precompact subset $F_l$ of $\R^k$; the
maps $\xi_l$ are also required to satisfy a certain smooth
compatibility condition, but we need not consider it here. The pairs
$(G_l,\xi_l)$ are called the {\em charts} of $\Theta$.

In order to cover this case, we need to make the following modifications
in the statement and in the proof of Theorem~\ref{thm:wu}. First of
all, let $\{P_\theta : \theta \in \Theta\}$ satisfy Condition 3) of the theorem, and replace Condition 2) with
\begin{enumerate}
\item[2a)] For each $l$, the map $u \mapsto P_{\xi^{-1}_l(u)}$, $u \in
  F_l$, is uniformly locally Lipschitz: there exist constants $r_l > 0$ and $m_l > 0$, such that for every $u \in F_l$,
$$
d_V(P_{\xi^{-1}_l(u)},P_{\xi^{-1}_l(w)}) \le m_l \| u - w \|
$$
for all $w \in B_{r_l} (u)$.
\end{enumerate}
[Note that $\xi^{-1}_l(u) \in G_l \subset \Theta$ for all $u \in
F_l$.] Condition 1) is satisfied for each $F_l$ by definition of
$\Theta$. Next, we need to modify the first-stage encoder. For each $l$,
quantize $F_l$ in cubes of side $1/\lceil n^{1/2} \rceil$, so that
each $u \in F_l$ can be encoded into $k(\log n^{1/2} + \log J_l)$ bits,
for some $J_l$, and reproduced by some $\wh{u} \in F_l$ satisfying $\|
u - \wh{u} \| \le \sqrt{k/n}$. Then $\theta = \xi^{-1}_l(u)$ and
$\wh{\theta} = \xi^{-1}_l(\wh{u})$ both lie in $G_l \subset
\Theta$. Now, when the first-stage encoder computes the
minimum-distance estimate $\td{\theta}$ of the active source $\theta$,
it will prepend a fixed-length binary description of the index $l$
such that $\td{\theta} \in G_l$ to the binary description of the cube
in $\R^k$ containing $\td{u} = \xi_l(\td{\theta})$. Let $\wh{u}$ be
the reproduction of $\td{u}$ under the cubic quantizer for $F_l$. The per-letter rate of the
resulting two-stage code is
$$
R(C^{n,n}) = R + O\left(\frac{\log n}{n}\right)
$$
bits per letter. The $n$th-order distortion redundancy is bounded as
$$
\delta^n_\theta(C^{n,n}) \le 4\rho_{\max} \left\{\E_\theta[
  d_V(P_\theta,P_{\td{\theta}})] + \E_\theta
  [d_V(P_{\td{\theta}},P_{\wh{\theta}})]\right\},
$$
where $\wh{\theta} = \xi^{-1}_l(\wh{u})$. The first term in the brackets is upper-bounded by means of the
usual Vapnik--Chervonenkis estimate,
$$
\E_\theta \left[ d_V(P_\theta,P_{\td{\theta}}) \right] \le 2c_1\sqrt{\frac{\log n}{n}}
+ \frac{3}{2n},
$$
while the second term is handled using Condition 2a). Specifically,
if $\td{\theta} \in G_l$, then $P_{\td{\theta}} =
P_{\xi^{-1}_l(\td{u})}$ and $P_{\wh{\theta}} =
P_{\xi^{-1}_l(\wh{u})}$. Then the same argument as in the proof of
Theorem~\ref{thm:wu} can be used to show that there exists a constant
$b_l > 0$ such that $d_V(P_{\td{\theta}},P_{\wh{\theta}}) \le b_l
\sqrt{k/n}$, which can be further bounded by $b\sqrt{k/n}$ with $b
\deq \max_l b_l$. Combining all these bounds and using Proposition~\ref{prop:drf}, we get that the distortion redundancy is
$$
\delta_\theta(C^{n,n}) = O\left(\sqrt{\frac{\log n}{n}}\right).
$$
This establishes that $\{C^{n,n}\}$ is weakly
minimax universal for the curved parametric family $\{P_\theta :
\theta \in \Theta\}$. The fidelity of the source identification procedure is similar to that in the "flat" case $\Theta \subset \R^k$, by the same Borel--Cantelli arguments as in the proof of Theorem~\ref{thm:wu}.

\subsection{Extension to unbounded distortion measures}
\label{ssec:ubound}

In this section we show that the boundedness condition on the
distortion measure can be relaxed, so that our approach can work with any distortion measure satisfying a certain moment condition, except
that the distortion redundancy will converge to zero at a slower rate of $O(\sqrt[4]{n^{-1}\log n})$ instead of $O(\sqrt{n^{-1}\log n})$, as in the bounded case.

Specifically, let $\{P_\theta
: \theta \in \Theta\}$ be a family of i.i.d. sources satysfing the conditions of Theorem~\ref{thm:wu}, and let $\map{\rho}{\cX
  \times \hcX}{\R^+}$ be a single-letter distortion function for which
there exists a {\em reference letter} $a_* \in \hcX$ such that
\begin{equation}
\int_\cX \rho^2(x,a_*) dP_\theta(x) \le G < \infty
\label{eq:refletter}
\end{equation}
for all $\theta \in \Theta$, and which has the form $\rho(x,\wh{x}) = [d(x,\wh{x})]^p$ for some $p$, where $d(\cdot,\cdot)$ is a metric on $\cX \cup \hcX$. In the following, we shall show that for any rate $R > 0$ satisfying
$$
D(R,\Theta) \deq \sup_{\theta \in \Theta} D_\theta(R) < \infty,
$$
and for any $\epsilon > 0$ there exists a sequence $\{C^{n,n}\}^\infty_{n=1}$ of two-stage $(n,n)$-block codes, such that
\begin{equation}
R(C^{n,n}) \le R + \epsilon + O\left(\frac{\log n}{n}\right)
\label{eq:rate_ub}
\end{equation}
and
\begin{equation}
\delta_\theta(C^{n,n}) \le \epsilon + O\left(\sqrt[4]{\frac{\log n}{n}}\right)
\label{eq:dist_ub}
\end{equation}
for every $\theta \in \Theta$. Taking a cue from Garc\'ia-Mu\~noz and Neuhoff \cite{GarNeu82}, we shall call a sequence of codes $\{C^{n,n}\}$ satisfying 
$$
\lim_{n \to \infty} R(C^{n,n}) \le R+\epsilon
$$
and
$$
\lim_{n \to \infty} \delta_\theta(C^{n,n}) \le \epsilon, \qquad \forall \theta \in \Theta
$$
for a given $\epsilon > 0$ {\em $\epsilon$-weakly minimax universal} for $\{P_\theta : \theta \in \Theta\}$. By continuity, the existence of $\epsilon$-weakly minimax universal codes for all $\epsilon > 0$ then implies the existence of weakly minimax universal codes in the sense of Definition~\ref{def:wmu}. Moreover, we shall show that the convergence rate of the source identification procedure is the same as in the case of a bounded distortion function, namely $O(\sqrt{n^{-1}\log n})$; in particular, the constant implicit in the $O(\cdot)$ notation depends neither on $\epsilon$ nor on the behavior of $\rho$.

The proof below draws upon some ideas of Dobrushin \cite{Dob70}, the
difference being that he considered {\em robust}, rather than universal, codes.\footnote{A sequence of lossy codes is (strongly) robust for a given class of information sources at rate $R$ (see, e.g., \cite{Sak69,Sak75,NeuGar87}) if its asymptotic performance on each source in the class is no worse than the supremum of the distortion-rate functions of all the sources in the class at $R$. Neuhoff and Garc\'ia-Mu\~noz \cite{NeuGar87} have shown that strongly robust codes occur more widely than strongly minimax universal codes, but less widely than weakly minimax universal ones.} Let $M > 0$ be a constant to be specified later, and define a single-letter distortion
function $\map{\rho_M}{\cX \times \hcX}{\R^+}$ by
$$
\rho_M(x,\wh{x}) \deq \left\{
\begin{array}{ll}
\rho(x,\wh{x}), & \mbox{ if } \rho(x,\wh{x}) \le M \\
M, &\mbox { if } \rho(x,\wh{x}) > M
\end{array}\right..
$$
Let $\bar{D}_\theta(C^{n,m})$ denote the average per-letter $\rho_M$-distortion
of an $(n,m)$-block code $C^{n,m}$ with respect to $P_\theta$, and
let $\bar{D}_\theta(R)$ denote the corresponding Shannon DRF. Then Theorem~\ref{thm:wu} guarantees that for every $R > 0$ there exists a  weakly minimax universal sequence $\{\bar{C}^{n,n}\}^\infty_{n=1}$ of two-stage $(n,n)$-block codes, such that
\begin{equation}
R(\bar{C}^{n,n}) = R + O\left(\frac{\log n}{n}\right)
\label{eq:barrate}
\end{equation}
and
\begin{equation}
\bar{D}_\theta(\bar{C}^{n,n}) = \bar{D}_\theta(R) + O\left(\sqrt{\frac{\log n}{n}}\right)
\label{eq:bardist}
\end{equation}
for all $\theta \in \Theta$.

We shall now modify $\bar{C}^{n,n}$ to obtain a new code $C^{n,n}$. Fix some $\delta > 0$, to be chosen later. Let $\{\bar{C}^n_{\td{s}} : \td{s} \in \td{\cS}\}$ be the collection of the second-stage codes of $\bar{C}^{n,n}$. Fix $\td{s} \in \td{\cS}$ and let $\bar{\Gamma}_{\td{s}}$ be the reproduction codebook of $\bar{C}^n_{\td{s}}$. Let $\Gamma_{\td{s}} \subset \hcX^n$ be the set consisting of (a) all codevectors in $\bar{\Gamma}_{\td{s}}$, (b) all vectors obtained by replacing $\lfloor \delta n \rfloor$ or fewer components of each codevector in $\bar{\Gamma}_{\td{s}}$ with $a_*$, and (c) the vector $a^n_*$. The size of $\Gamma_{\td{s}}$ can be estimated by means of Stirling's formula as
$$
|\Gamma_{\td{s}}| = |\bar{\Gamma}_{\td{s}}|\sum^{\lfloor \delta n \rfloor}_{i=0} {n \choose i} + 1 \le |\bar{\Gamma}_{\td{s}}|2^{n[h(\delta) + o(1)]} + 1,
$$
where $h(\delta) \deq -\delta \log \delta - (1-\delta) \log (1-\delta)$ is the binary entropy function. Since $|\bar{\Gamma}_{\td{s}}| = 2^{nR}$, we can choose $\delta$ small enough so that
\begin{equation}
|\Gamma_{\td{s}}| \le 2^{n(R+\epsilon)}.
\label{eq:newrate}
\end{equation}
Now, if $\bar{C}^n_{\td{s}}$ maps a given $x^n \in \cX^n$ to $\wh{\bar{x}}^n = (\wh{\bar{x}}_1,\cdots,\wh{\bar{x}}_n) \in \hcX^n$, define a new string $\wh{x}^n = (\wh{x}_1,\cdots,\wh{x}_n) \in \hcX^n$ as follows. If $|\{ 1 \le i \le n : \rho(x_i,\wh{\bar{x}}_i) > M \}| \le \delta n$, let
$$
\wh{x}_i \deq \left\{
\begin{array}{ll}
\wh{\bar{x}}_i, & \mbox{ if } \rho(x_i,\wh{\bar{x}}_i) \le M \\
a_*, & \mbox{ if } \rho(x_i,\wh{\bar{x}}_i) > M
\end{array}\right.;
$$
otherwise, let $\wh{x}^n = a^n_*$. Now, construct a new code $C^n_{\td{s}}$ with the codebook $\Gamma_{\td{s}}$, and with the encoder and the decoder defined in such a way that $C^n_{\td{s}}(x^n) = \wh{x}^n$ whenever $\bar{C}^n_{\td{s}}(x^n) = \wh{\bar{x}}^n$. Finally, let $C^{n,n}$ be a two-stage code with the same first-stage encoder as $\bar{C}^{n,n}$, but with the collection of the second-stage codes replaced by $\{C^n_{\td{s}} \}$. From (\ref{eq:newrate}) it follows that $R(C^{n,n}) \le R(\bar{C}^{n,n}) + \epsilon$. Since $R(\bar{C}^{n,n}) = R + O(n^{-1}\log n)$, we have that
\begin{equation}
R(C^{n,n}) \le R + \epsilon + O\left(\frac{\log n}{n}\right).
\label{eq:newrate_2}
\end{equation}
Furthermore, the code $C^{n,n}$ has the following property:

\begin{lemma} Let $G' = G(1+2/\delta)$. Then for any $\theta\in \Theta$,
\begin{equation}
D_\theta(C^{n,n}) \le \bar{D}_\theta(\bar{C}^{n,n}) + \sqrt{\frac{G'\bar{D}_\theta(\bar{C}^{n,n})}{M}}.
\label{eq:nobardist}
\end{equation}
\label{lm:barmismatch}
\end{lemma}

\begin{proof} See Appendix~\ref{app:barmismatch_proof}. \end{proof}

Substituting (\ref{eq:bardist}) into (\ref{eq:nobardist}), we have that
\begin{equation}
D_\theta(C^{n,n}) \le \bar{D}_\theta(R) + O\left(\sqrt{\frac{\log n}{n}}\right) + \sqrt{\frac{G'}{M} \bar{D}_\theta(R) + O\left(\sqrt{\frac{\log n}{n}}\right)}.
\label{eq:nobardist_2}
\end{equation}
Now, since $\rho_M(x,\wh{x}) \le \rho(x,\wh{x})$ for all $(x,\wh{x}) \in \cX \times \hcX$, $\bar{D}_\theta(R) \le D_\theta(R)$ for all $\theta \in \Theta$. Using this fact and the inequality $\sqrt{a+b} \le \sqrt{a} + \sqrt{b}$, we can write
$$
D_\theta(C^{n,n}) \le D_\theta(R) + O\left(\sqrt{\frac{\log n}{n}}\right)  + \sqrt{\frac{G'}{M}D_\theta(R)} + O\left(\sqrt[4]{\frac{\log n}{n}}\right).
$$
Upon choosing $M$ so that $\sqrt{G'D(R,\Theta)/M} < \epsilon$, we get
\begin{equation}
\delta_\theta(C^{n,n}) \le \epsilon + O\left(\sqrt[4]{\frac{\log n}{n}}\right).
\label{eq:newred}
\end{equation}
Thus, (\ref{eq:newrate_2}) and (\ref{eq:newred}) prove the claim made at the beginning of the section. Moreover, because the first-stage encoder of $C^{n,n}$ is the same as in $\bar{C}^{n,n}$, our code modification procedure has no effect on parameter estimation,
so the same arguments as in the end of the proof of
Theorem~\ref{thm:wu} can be used to show that the decoder can identify the source in effect up to a variational ball of radius $O(\sqrt{n^{-1}\log n})$ asymptotically almost surely.

\section{Examples}
\label{sec:examples}

In this section we present a detailed analysis of two classes of
parametric sources that meet Conditions 1)--3) of Theorem~\ref{thm:wu}, and thus admit schemes for joint universal lossy coding and modeling. These are finite mixture classes and exponential
families, which are widely used in statistical modeling, both
in theory and in practice (see, e.g., \cite{Sorals71,BarShe91,ZHPZ96,FigJai02}).

\subsection{Mixture classes}

Let $p_1,\cdots,p_k$ be fixed probability densities on a measurable $\cX \subseteq
\R^d$, and let
$$
\Theta \deq \left\{ \theta = (\theta_1,\cdots,\theta_k) \in \R^k: 0
\le \theta_i \le 1,  1 \le i \le k; \sum^k_{i=1} \theta_i = 1 \right\}
$$
be the probability $k$-simplex. Then the mixture class defined
by the $p_i$'s consists of all densities of the form
$$
p_\theta(x) = \sum^k_{i=1} \theta_i p_i(x).
$$
The parameter space $\Theta$ is obviously compact, which establishes
Condition 1) of Theorem~\ref{thm:wu}. In order to show that Condition
2) holds, fix any $\theta,\eta \in \Theta$. Then
\begin{eqnarray*}
d_V(P_\theta,P_\eta) &=& \frac{1}{2}\int_\cX |p_\theta(x) -
p_\eta(x)|dx \\
&\le& \frac{1}{2}\sum^k_{i=1}\int_\cX |\theta_i - \eta_i| p_i(x)dx \\
&=& \frac{1}{2}\sum^k_{i=1}|\theta_i - \eta_i| \\
&\le& \frac{\sqrt{k}}{2} \sqrt{\sum^k_{i=1} (\theta_i - \eta_i)^2} \\
&=& \frac{\sqrt{k}}{2} \|\theta - \eta\|,
\end{eqnarray*}
where the last inequality is a consequence of the concavity of the square root. This implies that the map $\theta \mapsto P_\theta$ is everywhere Lipschitz with
Lipschitz constant $\sqrt{k}/2$. We have left to show that Condition 3) of
Theorem~\ref{thm:wu} holds as well, i.e., that the Yatracos class
$$
\cA_\Theta = \Big\{ A_{\theta,\eta} = \{x \in \cX: p_\theta(x) >
p_\eta(x) \} : \theta,\eta \in \Theta; \theta \neq \eta \Big\}
$$
has finite Vapnik--Chervonenkis dimension. To this end, observe that
$x \in A_{\theta,\eta}$ if and only if
$$
\sum^k_{i=1} (\theta_i - \eta_i)p_i(x) > 0.
$$
Thus, $\cA_\Theta$ consists of sets of the form
$$
\left\{ x \in \cX: \sum^k_{i=1} \alpha_i p_i(x) > 0, \alpha =
(\alpha_1,\cdots,\alpha_k) \in \R^k \right\}.
$$
Since the functions $p_1,\cdots,p_k$ span a linear space whose dimension is not larger than $k$, Lemma~\ref{lm:vc_linspace} in the Appendices guarantees that $\sV(\cA_\Theta)
\le k$, which establishes Condition 3).

\subsection{Exponential families}

Let $\cX$ be a measurable subset of $\R^d$, and let $\Theta$ be a
compact subset of $\R^k$. A family $\{ p_\theta : \theta \in \Theta\}$
of probability densities on $\cX$ is an {\em exponential family} \cite{BarShe91,AmaNag00} if
each $p_\theta$ has the form
\begin{eqnarray}
p_\theta(x) &=& p(x) \exp \left( \sum^k_{i=1}\theta_i h_i(x) -
g(\theta) \right) \nonumber\\
&\equiv& p(x) e^{\theta \cdot h(x) -
g(\theta)},
\label{eq:expfam}
\end{eqnarray}
where $p$ is a fixed reference density, $h_1,\cdots,h_k$ are
fixed real-valued functions on $\cX$, and
$$
g(\theta) = \ln \int_\cX e^{\theta \cdot h(x)}p(x)dx
$$
is the normalization constant. By way of notation, $h(x) \deq
(h_1(x),\cdots,h_k(x))$ and $\theta \cdot h(x) \deq \sum^k_{i=1}
\theta_i h_i(x)$. Given the densities $p$ and $p_\theta$, let $P$ and
$P_\theta$ denote the corresponding distributions. The assumed
compactness of $\Theta$ guarantees that the family $\{P_\theta : \theta \in
\Theta\}$ satisfies Condition 1) of
Theorem~\ref{thm:wu}. In the following, we shall demonstrate that Conditions 2)
and 3) can also be met under certain regularity assumptions.

It is customary to choose the functions $h_i$ in such a way that $\{1,h_1,\cdots,h_k\}$ is a linearly
independent set. This guarantees that the map $\theta \mapsto
P_\theta$ is one-to-one. We shall also assume that each $h_i$ is
square-integrable with respect to $P$:
$$
\int_\cX h_i^2 dP \equiv \int_\cX h^2_i(x) p(x) dx < \infty, \qquad 1 \le i
\le k.
$$
Then the $(k+1)$-dimensional real linear space $\cF \subset L^2(\cX,P)$ spanned by $\{1,h_1,\cdots,h_k\}$ can
be equipped with an inner product
$$
\ave{f,g} \deq \int_\cX fg dP, \qquad f,g \in \cF
$$
and the corresponding $L_2$ norm
$$
\| f \|_2 \deq \sqrt{ \ave{f,f} } \equiv \sqrt{\int_\cX f^2 dP},
\qquad f \in \cF.
$$
Also let
$$
\| f \|_\infty \deq \inf \big\{ M : |f(x)| \le M \mbox{ $P$-a.e.} \big\}
$$
denote the $L_\infty$ norm of $f$. Since $\cF$ is
finite-dimensional, there exists a constant $A_k > 0$ such that
$$
\| f \|_\infty \le A_k \| f \|_2.
$$
Finally, assume that the logarithms of Radon--Nikodym derivatives $dP/dP_\theta \equiv p/p_\theta$ are uniformly bounded $P$-a.e.: $\Sup_{\theta \in \Theta} \| \log p/p_\theta \|_\infty < \infty$. Let
$$
D(P_\theta \| P_\eta) \deq \int_\cX \frac{dP_\theta}{dP_\eta} \ln \frac{dP_\theta}{dP_\eta} dP_\eta \equiv \int_\cX p_\theta \ln
\frac{p_\theta}{p_\eta} dx
$$
denote the relative entropy (information divergence) between $P_\theta$ and $P_\eta$. Then we have the following basic estimate:

\begin{lemma} For all $\theta,\eta \in \Theta$,
$$
D(P_\theta \| P_\eta) \le \frac{1}{2}e^{\| \ln p/p_\theta
  \|_\infty} e^{2 A_k \| \theta - \eta \|} \| \theta - \eta
  \|^2,
$$
where $\| \cdot \|$ is the Euclidean norm on $\R^k$.
\end{lemma}

\begin{proof} The proof is along the lines of Barron and Sheu
  \cite[Lemma~5]{BarShe91}. Without loss of generality, we may assume that the
  functions $\{h_0,h_1,\cdots,h_k\}$, $h_0 \equiv 1$, form an
  orthonormal set with respect to $P$:
$$
\ave{h_i,h_j} = \int_\cX h_ih_jdP = \delta_{ij}, \qquad 0 \le i,j \le
k.
$$
Then
$$
\| (\theta - \eta) \cdot h \|_2 = \left\| \sum^k_{i=1} (\theta_i -
\eta_i) h_i \right\|_2 = \| \theta - \eta \|.
$$
Now, since
$$
g(\eta) - g(\theta) = \ln \int_\cX e^{(\eta - \theta) \cdot
  h}dP_\theta = \ln \int_\cX e^{(\eta - \theta) \cdot h(x)}
p_\theta(x)dx,
$$
we have
$$
| g(\eta) - g(\theta) | \le \| (\eta - \theta) \cdot h
\|_\infty \le A_k \| (\eta - \theta) \cdot h \|_2 \\ = A_k \| \eta -
\theta \|.
$$
Furthermore,
$$
\ln \frac{p_\theta}{p_\eta} = (\theta - \eta) \cdot h + g(\eta) -
g(\theta),
$$
whence it follows that the logarithm of the Radon--Nikodym derivative
$dP_\theta/dP_\eta = p_\theta/p_\eta$ is bounded $P$-a.e.: $\| \ln p_\theta/p_\eta \|_\infty \le 2 \|
(\eta - \theta) \cdot h \|_\infty \le 2A_k \| \eta - \theta \|$. In
this case, the relative entropy $D(P_\theta \| P_\eta)$ satisfies
\cite[Lemma~1]{BarShe91}
$$
D(P_\theta \| P_\eta) \le \frac{1}{2} e^{\| \ln p_\theta/p_\eta -
  c\|_\infty} \int_\cX \left( \ln \frac{p_\theta}{p_\eta} -
c\right)^2 dP_\theta
$$
for any constant $c$. Choosing $c = g(\eta) - g(\theta)$ and using the
orthonormality of the $h_i$, we get
\begin{eqnarray*}
D(P_\theta \| P_\eta) &\le& \frac{1}{2} e^{\| (\theta - \eta) \cdot
  h\|_\infty} \int_\cX ((\theta - \eta) \cdot h)^2 dP_\theta\\
&=& \frac{1}{2} e^{ \| (\theta - \eta) \cdot h \|_\infty} \int_\cX
\frac{p_\theta}{p} \left( (\theta - \eta) \cdot h \right)^2 dP\\
&\le& \frac{1}{2} e^{\| \ln p/p_\theta \|_\infty} e^{2A_k \| \theta -
  \eta \|} \| (\theta - \eta) \cdot h \|^2_2\\
&=& \frac{1}{2} e^{\| \ln p/p_\theta \|_\infty} e^{2A_k \| \theta -
  \eta \|} \| \theta - \eta \|^2,
\end{eqnarray*}
and the lemma is proved.
\end{proof}

Now, using Pinsker's inequality $d_V(P_\theta,P_\eta) \le
\sqrt{(1/2)D(P_\theta \| P_\eta)}$ \cite[Lemma~5.2.8]{Gra90a} together with the above lemma and the assumed uniform boundedness of $\ln p/p_\theta$, we get the bound
\begin{equation}
d_V(P_\theta,P_\eta) \le m_0 e^{A_k\|\theta - \eta \|} \| \theta - \eta \|, \qquad
\theta,\eta \in \Theta,
\label{eq:dvbound}
\end{equation}
where $m_0 \deq \frac{1}{2}\exp{\left(\frac{1}{2} \Sup_{\theta \in \Theta} \| \ln p/p_\theta \|_\infty \right)}$. If we fix $\theta \in \Theta$, then from (\ref{eq:dvbound}) it
follows that, for any $r > 0$,
$$
d_V(P_\theta,P_\eta) \le m_0 e^{A_k r} \| \theta - \eta \|
$$
for all $\eta$ satisfying $\| \eta - \theta \| \le r$. That is,
the family $\{P_\theta : \theta \in \Theta \}$ satisfies the uniform local
Lipschitz condition [Condition 2) of Theorem~\ref{thm:wu}], and the magnitude of the Lipschitz constant can be controlled
by tuning $r$.

All we have left to show is that the Vapnik--Chervonenkis condition
[Condition 3) of Theorem~\ref{thm:wu}] is satisfied. Let $\theta, \eta
\in \Theta$ be distinct; then $p_\theta(x) > p_\eta(x)$ if and only if
$(\theta - \eta) \cdot h(x) > g(\theta) - g(\eta)$. Thus, the
corresponding Yatracos
class $\cA_\Theta$ consists of sets of the form
$$
\left\{x \in \cX : \alpha_0 + \sum^k_{i=1} \alpha_i h_i(x) > 0, \alpha
= (\alpha_0,\alpha_1,\cdots,\alpha_k) \in \R^{k+1}\right\}.
$$
Since the functions $1,h_1,\cdots,h_k$ span a $(k+1)$-dimensional
linear space, $\sV(\cA_\Theta) \le k+1$ by
Lemma~\ref{lm:vc_linspace}. 

\section{Summary and discussion}
\label{sec:summary}

We have constructed and analyzed a scheme for universal fixed-rate lossy coding of continuous-alphabet i.i.d. sources based on a forward relation between statistical modeling and universal coding, in the spirit of Rissanen's achievability theorem \cite[Theorem~1b]{Ris84} (see also Theorem~2 in \cite{ChoEffGra96}). To the best of our knowledge, such a joint universal source coding and source modeling scheme has not been constructed before, although Chou {\em et al.} \cite{ChoEffGra96} have demonstrated the existence of universal vector quantizers whose Lagrangian redundancies converge to zero at the same rate as the corresponding redundancies in Rissanen's achievability theorem for the lossless case. What we have shown is that, for a wide class of bounded distortion measures and for any compactly parametrized family of i.i.d.~sources with absolutely continuous distributions satisfying a smoothness condition and a Vapnik--Chervonenkis learnability condition, the tasks of parameter estimation (statistical modeling) and universal lossy coding can be accomplished jointly in a two-stage set-up, with the overhead per-letter rate and the distortion redundancy converging to zero as $O(n^{-1}\log n)$ and $O(\sqrt{n^{-1}\log n})$, respectively, as the block length $n$ tends to infinity, and the extra bits generated by the first-stage encoder can be used to identify the active source up to a variational ball of radius $O(\sqrt{n^{-1}\log n})$ (a.s.). We have compared our scheme with several existing schemes for universal vector quantization and demonstrated that our approach offers essentially similar performance in terms of rate and distortion, while also allowing the decoder to reconstruct the statistics of the source with good precision. We have described an extension of our scheme to unbounded distortion measures satisfying a moment condition with respect to a reference letter, which suffers no change in overhead rate or in source estimation fidelity, although it gives a slower, $O(\sqrt[4]{n^{-1}\log n})$, convergence rate for distortion redundancies. Finally, we have presented detailed examples of parametric sources satisfying the conditions of our Theorem~\ref{thm:wu} (namely, finite mixture classes and exponential families) and thus admitting schemes for joint universal quantization and modeling.

As mentioned in the Introduction, in treating universal lossy source coding as a statistical problem the term ``model" can refer either to a probabilistic description of the source or to a probabilistic description of a rate-distortion codebook. In fact, as shown by Kontoyiannis and Zhang \cite{KonZha02}, for variable-rate lossy codes operating under a fixed distortion constraint, there is a one-to-one correspondence between codes and discrete distributions over sequences in the reproduction space (satisfying suitable ``admissibility" conditions), which they dubbed the ``lossy Kraft inequality." The same paper also demonstrated the existence of variable-rate universal lossy codes for finite-alphabet memoryless sources with rate redundancy converging to zero as $(k/2)\log n/n$, where $k$ is the dimension of the simplex of probability distributions on the reproduction alphabet. Yang and Zhang \cite{YanZha02} proved an analogous result for fixed-rate universal lossy codes and showed furthermore that the $(k/2)\log n/n$ convergence rate is {\em optimal} in a certain sense. (The redundancies in our scheme are therefore suboptimal, as can be seen from comparing them to these bounds, as well as to those of Chou {\em et al.} \cite{ChoEffGra96}. It is certainly an interesting open problem to determine lower bounds on the redundancies in the setting of joint source coding and identification.) These papers, together with the work of Madiman, Harrison and Kontoyiannis \cite{MadKon04,MadHarKon04}, can be thought of as generalizing Rissanen's MDL principle to lossy setting, provided that the term ``model" is understood to refer to probability distributions over codebooks in the reproduction space.

We close by outlining several potential directions for further research. First of all, it would be of both theoretical and practical interest to extend the results presented here to sources with memory in order to allow more realistic source models such as autoregressive or Markov sources, and to variable-rate codes, so that unbounded parameter spaces could be accommodated. We have made some initial progress in this direction in \cite{Rag06a,Rag07}, where we constructed joint schemes for variable-rate universal lossy coding and identification of stationary ergodic sources satisfying a certain mixing condition. Moreover, the theory presented here needs to be tested in practical settings, one promising area for applications being {\em media forensics} \cite{MouKoe05}, where the parameter $\theta$ could represent traces or ``evidence" of some prior processing performed, say, on an image or on a video sequence, and where the goal is to design an efficient system for compressing the data for the purposes of transmission or storage in such a way that the evidence can be later recovered from the compressed signal with minimal degradation in fidelity.

\section*{Acknowledgment}

The author would like to thank Andrew R.~Barron, Ioannis Kontoyiannis, Mokshay Madiman and Pierre Moulin for useful discussions. Insightful comments and advice by the anonymous reviewers and the Associate Editor Michelle Effros, which significantly helped improve the presentation, are also gratefully acknowledged.

\useRomanappendicesfalse
\appendices
\renewcommand{\theequation}{\thesection.\arabic{equation}}
\setcounter{equation}{0}

\renewcommand{\thedefinition}{\Alph{section}.\arabic{definition}}
\renewcommand{\theremark}{\Alph{section}.\arabic{remark}}
\renewcommand{\thelemma}{\Alph{section}.\arabic{lemma}}
\renewcommand{\thetheorem}{\Alph{section}.\arabic{theorem}}

\section{Vapnik--Chervonenkis theory}
\label{app:vc}

In this appendix, we summarize, for the reader's convenience, some basic
concepts and results of the Vapnik--Chervonenkis theory. A detailed
treatment can be found, e.g., in \cite{DevLug01}.

\begin{definition}[shatter coefficient]
\label{def:shatter}
 Let $\cA$ be an arbitrary collection of measurable
  subsets of $\R^d$. Given an $n$-tuple $x^n = (x_1,\cdots,x_n) \in
  (\R^d)^n$, let $\cA(x^n)$ be the subset of $\{0,1\}^n$ obtained by
  listing all distinct binary strings of the form $(1_{\{x_1 \in
    A\}},\cdots,1_{\{x_n \in A\}})$ as $A$ is varied over $\cA$. Then
$$
\sS_\cA(n) \deq \max_{x^n \in (\R^d)^n} |\cA(x^n)|
$$
is called the {\em $n$th shatter coefficient} of $\cA$.
\end{definition}

\begin{definition}[VC dimension; VC class]
\label{def:vc}
The largest integer $n$ for which $\sS_\cA(n) =
  2^n$ is called the {\em Vapnik--Chervonenkis dimension} (or the VC
  dimension, for short) of $\cA$ and denoted by $\sV(\cA)$. If
  $\sS_\cA(n) = 2^n$ for all $n=1,2,\cdots$, then we define $\sV(\cA)
  = \infty$. If $\sV(\cA) < \infty$, we say that $\cA$ is a {\em
    Vapnik--Chervonenkis class} (or VC class).
\end{definition}

The basic result of Vapnik--Chervonenkis theory relates the shatter
coefficient $\sS_\cA(n)$ to uniform
deviations of the probabities of events in $\cA$ from their relative
frequencies with respect to an i.i.d. sample of size $n$:

\begin{lemma}[the Vapnik--Chervonenkis inequalities]
\label{lm:vc}
Let $\cA$ be an arbitrary collection of measurable subsets of $\R^d$,
and let $X^n = (X_1,\ldots,X_n)$ be an $n$-tuple of i.i.d. random
variables in $\R^d$ with the common distribution $P$. Then
\begin{equation}
\Pr \left\{ \sup_{A \in \cA}|P_{X^n}(A) - P(A)| > \epsilon \right\}
\le 8 \sS_\cA(n) e^{-n\epsilon^2/32}
\label{eq:vcbound1}
\end{equation}
for any $\epsilon > 0$, and
\begin{equation}
\E \left\{ \sup_{A \in \cA} |P_{X^n}(A) - P(A)| \right\} \le 2
\sqrt{\frac{\log 2\sS_\cA(n)}{n}},
\label{eq:vcbound2}
\end{equation}
where $P_{X^n}$ is the empirical distribution of $X^n$:
$$
P_{X^n}(B) \deq \frac{1}{n} \sum^n_{i=1} 1_{X_i \in B}
$$
for all Borel sets $B \subset \R^d$. The probabilities and
expectations are with respect to $P$.
\end{lemma}

Now, if $\cA$ is a VC class and $\sV(\cA) \ge 2$, then the results of
Vapnik and Chervonenkis \cite{VC71} and Sauer \cite{Sau72} imply that
$\sS_\cA(n) \le n^{\sV(\cA)}$. Plugging this bound into (\ref{eq:vcbound1}) and (\ref{eq:vcbound2}), we obtain the
following:

\begin{lemma}\label{lm:vcclass_bound}
If $\cA$ is a VC class with $\sV(\cA) \ge 2$, then
\begin{equation}
\Pr \left\{ \sup_{A \in \cA} |P_{X^n}(A) - P(A)| > \epsilon \right\}
\le 8n^{\sV(\cA)}e^{-n\epsilon^2/32}
\label{eq:vcclass_bound1}
\end{equation}
for any $\epsilon > 0$, and
\begin{equation}
\E \left\{ \sup_{A \in \cA} |P_{X^n}(A) - P(A)| \right\} \le c
\sqrt{\frac{\log n}{n}},
\label{eq:vcclass_bound2}
\end{equation}
where $c$ is a constant that depends only on $\sV(\cA)$.
\end{lemma}

\begin{remark} One can use more delicate arguments involving metric
  entropies and covering numbers, along the lines of Dudley
  \cite{Dud78}, to improve the bound in (\ref{eq:vcclass_bound2})
  to $c'\sqrt{1/n}$, where $c' = c'(\sV(\cA))$ is another
  constant. However, $c'$ turns out to be much larger than $c$, so that, for all
  ``practical'' values of $n$, the "improved" $O(\sqrt{1/n})$ bound is
  much worse than the original $O(\sqrt{n^{-1}\log n})$ bound.
\end{remark}

\begin{lemma}\label{lm:vc_linspace}
Let $\cF$ be an $m$-dimensional linear space of
  real-valued functions on $\R^d$. Then the class
$$
\cA = \Big\{ \{x : f(x) \ge 0 \} : f \in \cF \Big\}
$$
is a VC class, and $\sV(\cA) \le m$.
\end{lemma}

\setcounter{equation}{0}

\section{Proof of Proposition~\ref{prop:drf}}
\label{app:drf}

Fix $\theta \in \Theta$, and let $X$ be distributed according to $P_\theta$. Let the distortion function $\rho$ satisfy Condition 1) of Proposition~\ref{prop:drf}. Then a result of \Csiszar\ \cite{Csi74} says that, for each point $(R,D_\theta(R))$ on the distortion-rate curve for $P_\theta$, there exists a random variable $Y$ with values in the reproduction alphabet $\hcX$, where the joint distribution of $X$ and $Y$ is such that
\begin{equation}
I(X,Y) = R \qquad {\rm and} \qquad \E_{XY}[\rho(X,Y)] = D_\theta(R),
\label{eq:csi1}
\end{equation}
and the Radon--Nikodym derivative
$$a(x,y) \deq \frac{dP_{XY}}{d(P_X \times P_Y)}(x,y),
$$
where $P_X \equiv P_\theta$, has the parametric form
\begin{equation}
a(x,y) = \alpha(x) 2^{-s\rho(x,y)},
\label{eq:csi2}
\end{equation}
where $s \ge 0$ and $\alpha(x) \ge 1$ satisfy
\begin{equation}
\int_\cX \alpha(x) 2^{-\rho(x,y)}dP_\theta(x) \le 1, \qquad \forall y \in \hcX,
\label{eq:csi3}
\end{equation}
and $-1/s = D'_\theta(R)$, the derivative of the DRF $D_\theta(R)$ at $R$, i.e., $-1/s$ is the slope of the tangent to the graph of $D_\theta(R)$ at $R$.

Next, let $N = \lfloor 2^{n(R+\delta)} \rfloor$, where $\delta > 0$ will be specified later, and generate a random codebook $\cW$ as a vector $\cW = (W_1,\cdots,W_N)$, where each $W_i = (W_{i1},\cdots,W_{in}) \in \hcX^n$, and the $W_{ij}$'s are i.i.d. according to $P_Y$. Thus,
$$
P_\cW = \bigtimes^N_{i=1} P^n_Y
$$
is the probability distribution for the randomly selected codebook. We also assume that $\cW$ is independent from $X^n = (X_1,\cdots,X_n)$. Now, let $C_\cW$ be a (random) $n$-block code with the reproduction codebook $\cW$ and the minimum-distortion encoder, so that $\rho(x^n,C_\cW(x^n)) = \rho(x^n,\cW)$, where $\rho(x^n,\cW) \deq \Min_{1 \le i \le N} \rho(x^n,W_i)$. Then the average per-letter distortion of this random code over the codebook generation and the source sequence is
\begin{eqnarray*}
\Delta_n &=& \int D_\theta(C_{\bd{w}}) dP_\cW(\bd{w}) \\
&=& \frac{1}{n}\int \E_\theta\left[\rho(X^n,\bd{w})\right] dP_\cW(\bd{w}) \\
&=& \frac{1}{n}\int \left(\int_{\cX^n} \rho(x^n,\bd{w})dP_\theta(x^n) \right) dP_\cW(\bd{w}).
\end{eqnarray*}
Using standard arguments (see, e.g., Gallager's proof of the source coding theorem \cite[Ch.~9]{Gal68}), we can bound $\Delta_n$ from above as
\begin{equation}
\Delta_n \le D_\theta(R) + \delta + \rho_{\max} \left(P_{XY}(Y^n \not\in \cS_{X^n}) + e^{-N2^{-n(R+\delta/2)}} \right),
\label{eq:delta_n_2}
\end{equation}
where
$$
\cS_{x^n} \deq \left\{ y^n \in \hcX^n : \rho(x^n,y^n) \le n(D_\theta(R)+\delta) \mbox{ and } i_n(x^n,y^n) \le n(R + \delta/2) \right\},
$$
and
$$
i_n(x^n,y^n) \deq \sum^n_{i=1} \log a(x_i,y_i) \equiv \log a(x^n,y^n).
$$
is the sample mutual information. Here, the pairs $(X_i,Y_i)$ are i.i.d. according to $P_{XY}$. Now, by the union bound,
\begin{equation}
P_{XY}(Y^n \not\in \cS_{X^n}) \le P_{XY} \left(\frac{1}{n} \sum^n_{i=1} \log a(X_i,Y_i) \ge R + \delta/2\right) + P_{XY} \left(\frac{1}{n} \sum^n_{i=1}\rho(X_i,Y_i) \ge D_\theta(R) + \delta\right).\label{eq:probs}
\end{equation}
Note that from (\ref{eq:csi1}) we have that $\E_{XY}[\log a(X,Y)] = I(X,Y) = R$ and $\E_{XY}[\rho(X,Y)] = D_\theta(R)$. Since $0 \le \rho(X,Y) \le \rho_{\max}$, the second probability on the right-hand side of (\ref{eq:probs}) can be bounded using Hoeffding's inequality \cite{Hoe63}, which states that for i.i.d. random variables $S_1,\cdots,S_n$ satisfying $a \le S_i \le b$ a.s.,
$$
\Pr \left(\frac{1}{n}\sum^n_{i=1}S_i \ge \E[S_1] + \delta \right) \le e^{-2n\delta^2/(b-a)^2}.
$$
This yields the estimate
\begin{equation}
P_{XY} \left(\frac{1}{n}\sum^n_{i=1}\rho(X_i,Y_i) \ge D_\theta(R) + \delta \right) \le e^{-2n\delta^2/\rho^2_{\max}}.
\label{eq:ldp2_2}
\end{equation}
In order to apply Hoeffding's inequality to the first probability on the right-hand side of (\ref{eq:probs}), we have to show that $\log a(X,Y)$ is bounded. From (\ref{eq:csi2}) we have that $\log a(x,y) = \log \alpha(x) - s\rho(x,y)$. On the other hand, integrating both sides of (\ref{eq:csi2}) with respect to $P_Y$, we get
$$
\alpha(x) = \frac{1}{\int 2^{-s\rho(x,y)}dP_Y(y)} \qquad P_\theta\mbox{-a.e.}
$$
Since $2^{-s\rho_{\max}} \le 2^{-s\rho(x,y)} \le 1$, we have that $1 \le \alpha(x) \le 2^{s\rho_{\max}}$, whence it follows that $-s\rho_{\max} \le \log a(x,y) \le s\rho_{\max}$. Thus, by Hoeffding's inequality,
\begin{equation}
P_{XY} \left( \frac{1}{n}\sum^n_{i=1} \log a(X_i,Y_i) \ge R + \delta/2 \right) \le e^{-n\delta^2/8s^2\rho^2_{\max}}.
\label{eq:ldp1_2}
\end{equation}
Putting together (\ref{eq:delta_n_2}), (\ref{eq:ldp2_2}) and (\ref{eq:ldp1_2}), and using the fact that $N \ge 2^{n(R + \delta)} - 1$, we obtain
$$
\Delta_n \le D_\theta(R) + \delta + \rho_{\max} \Big(e^{-2n\delta^2/\rho^2_{\max}} + e^{-n\delta^2/8s^2\rho^2_{\max}} + e^{-2^{n\delta/2}} \Big).
$$
Since $\Delta_n$ is the average of the expected distortion over the random choice of codes, it follows that there exists at least one code whose average distortion with respect to $P_\theta$ is smaller than $\Delta_n$. Thus,
$$
\wh{D}^n_\theta(R+\delta) \le D_\theta(R) + \delta + \rho_{\max} \left(e^{-2n\delta^2/\rho^2_{\max}} + e^{-n\delta^2/8s^2\rho^2_{\max}} + e^{-2^{n\delta/2}} \right).
$$
Now, let $c_s = \max(\rho_{\max}/2,2s\rho_{\max})$ and put $\delta = c_s \sqrt{n^{-1}\ln n}$ to get
\begin{equation}
\wh{D}^n_\theta\left(R+c_s\sqrt{\frac{\ln n}{n}}\right) - D_\theta(R) = (c_s + o(1))\sqrt{\frac{\ln n}{n}}.
\label{eq:drfbound}
\end{equation}
Because $-1/s$ is the slope of the tangent to the distortion-rate curve at the point $(R,D_\theta(R))$, and because $D_\theta(R)$ is nonincreasing in $R$, we have $-1/s' \le - 1/s$ for $s'$ corresponding to another point $(R',D_\theta(R'))$ with $R' < R$. Thus, $c_{s'} \le c_s$, and (\ref{eq:drfbound}) remains valid for all $R' \le R$. Thus, let $R' = R - c_s \sqrt{n^{-1}\ln n}$ to get
$$
\wh{D}^n_\theta(R) - D_\theta\left(R - c_s \sqrt{\frac{\ln n}{n}}\right) \le (c_s + o(1))\sqrt{\frac{\ln n}{n}}.
$$
Therefore, expanding $D_\theta(R)$ in a Taylor series to first order and recalling that $-1/s = D'_\theta(R)$, we see that
$$
\wh{D}^n_\theta(R) = c_s\left(1 + \frac{1}{s} + o(1)\right) \sqrt{\frac{\ln n}{n}},
$$
and the proposition is proved.

\section{Quantizer mistmatch lemma}
\label{app:mismatch}

\begin{lemma}
\label{lm:qmismatch}
Let $P$ and $Q$ be two absolutely continuous probability
  distributions on $\cX \subseteq \R^d$, with respective densities $p$
  and $q$, and let $\map{\rho}{\cX \times \hcX}{\R^+}$ be a single-letter distortion measure having the form $\rho(x,\wh{x}) = [d(x,\wh{x})]^p$, where $p > 0$ and $d(\cdot,\cdot)$ is a bounded metric on $\cX \cup \hcX$. Consider an $n$-block lossy code $C^n$ with the nearest-neighbor encoder,
  and let
$$
D_P(C^n) \deq \frac{1}{n} \E_P [\rho(X^n,C^n(X^n))] = \frac{1}{n}
\int_{\cX^n} \rho(x^n,C^n(x^n)) dP(x^n)
$$
be the average per-letter distortion of $C^n$ with respect to
$P$. Define $D_Q(C^n)$ similarly. Then
\begin{equation} \label{eq:mismatch_1}
|D_P(C^n)^{1/p} - D_Q(C^n)^{1/p}| \le 2^{1/p} d_{\max} d_V(P,Q).
\end{equation}
Furthermore, the corresponding $n$th-order operational DRF's $\wh{D}^n_P(R)$
and $\wh{D}^n_Q(R)$ satisfy
\begin{equation} \label{eq:mismatch_2}
|\wh{D}^n_P(R)^{1/p} - \wh{D}^n_Q(R)^{1/p}| \le 2^{1/p} d_{\max}d_V(P,Q).
\end{equation}
\end{lemma}

\begin{proof} The proof closely follows Gray, Neuhoff and Shields \cite{GraNeuShi75}. Let $\cP_n(P,Q)$ denote the set of all probability measures on $\cX^n \times \cX^n$ having $P^n$ and $Q^n$ as marginals, and let $\bar{\mu}$ achieve (or come arbitrarily close to) the infimum in the {\em Wasserstein metric}
$$
\bar{\rho}_n(P,Q) \deq  \inf_{\mu \in \cP_n(P,Q)} \left( \frac{1}{n} \int_{\cX^n \times \cX^n} \rho(x^n,y^n) d\mu(x^n,y^n) \right)^{1/p}.
$$
Suppose that $D_P(C^n) \le D_Q(C^n)$. Then, using the fact that $d$ is a metric, Minkowski's inequality, and the nearest-neighbor property of $C^n$, we have
\begin{eqnarray*}
D_P(C^n)^{1/p} &=& \left(\frac{1}{n} \int_{\cX^n} \rho(x^n,C^n(x^n)) dP^n(x^n)\right)^{1/p} \\
&=& \left(\frac{1}{n} \int_{\cX^n \times \cX^n} \rho(x^n,C^n(x^n)) d\bar{\mu}(x^n,y^n)\right)^{1/p} \\
&\le& \left(\frac{1}{n} \int_{\cX^n \times \cX^n} \rho(x^n,y^n) d\bar{\mu}(x^n,y^n) \right)^{1/p} + \left(\frac{1}{n} \int_{\cX^n \times \cX^n} \rho(y^n,C^n(y^n)) d\bar{\mu}(x^n,y^n) \right)^{1/p} \\
&=& \bar{\rho}_n(P,Q) + D_Q(C^n)^{1/p},
\end{eqnarray*}
Now,
$$
\bar{\rho}_n(P,Q) = \bar{\rho}_1(P,Q) = \left(\inf_{\mu \in \cP_1(P,Q)} \int_{\cX \times \cX} \rho(x,y)d\mu(x,y) \right)^{1/p}
$$
(see, e.g., \cite{GraNeuShi75}, Section~2), and $\rho(x,y) \le d^p_{\max} 1_{\{ x \neq y \}}$, so
$$
\inf_{\mu \in \cP_1(P,Q)} \int_{\cX \times \cX} \rho(x,y) d\mu(x,y) \le d^p_{\max} \inf_{\mu \in \cP_1(P,Q)} \int_{\cX \times \cX} 1_{\{ x \neq y \}} d\mu(x,y).
$$
The right-hand side of this expression is the well-known coupling characterization of twice the variational distance $d_V(P,Q)$ (see, e.g., Section~I.5 of Lindvall \cite{Lin02}), so we obtain
$$
D_P(C^n)^{1/p} \le D_Q(C^n)^{1/p} + 2^{1/p} d_{\max}d_V(P,Q).
$$
Interchanging the roles of $P$ and $Q$, we obtain (\ref{eq:mismatch_1}).

To prove (\ref{eq:mismatch_2}), let $C^n_*$
achieve the $n$th-order optimum for $P$: $D_P(C^n_*) =
\wh{D}^n_P(R)$. Without loss of generality, we can assume that $C^n_*$ has a nearest-neighbor encoder. Then
$$
\wh{D}^n_Q(R)^{1/p} \le D_Q(C^n_*)^{1/p} \le D_P(C^n_*)^{1/p} + 2^{1/p} d_{\max} d_V(P,Q) = \wh{D}^n_P(R)^{1/p} + 2^{1/p} d_{\max} d_V(P,Q).
$$
The other direction is proved similarly.
\end{proof}

\setcounter{equation}{0}

\section{Proof of Lemma~\ref{lm:barmismatch}}
\label{app:barmismatch_proof}

Fix a $\theta \in \Theta$. Define the measurable set $\cU \deq \{(x^n,z^n) \in \cX^n \times \cX^n :
C^{n,n}(x^n,z^n) = a^n_* \}$. Then the distortion $D_\theta(C^{n,n})$
can be split into two terms as
\begin{eqnarray}
D_\theta(C^{n,n}) &=& \frac{1}{n}\int_{\cX^n \times \cX^n} \rho(x^n,C^{n,n}(x^n,z^n))
dP_\theta(x^n,z^n) \nonumber \\
&=& \frac{1}{n}\int_\cU
\rho(x^n,C^{n,n}(x^n,z^n))dP_\theta(x^n,z^n) + \frac{1}{n} \int_{\cU^c}
\rho(x^n,C^{n,n}(x^n,z^n))dP_\theta(x^n,z^n),
\label{eq:barmismatch}
\end{eqnarray}
where the superscript $c$ denotes set-theoretic complement. We shall
prove the lemma by upper-bounding separately each of the two terms on the
right-hand side of (\ref{eq:barmismatch}).

First of all, we have
\begin{equation}
\frac{1}{n}\int_\cU \rho_M(x^n,\bar{C}^{n,n}(x^n,z^n))dP_\theta(x^n,z^n) \le
\frac{1}{n} \int_{\cX^n \times \cX^n}
\rho_M(x^n,\bar{C}^{n,n}(x^n,z^n))dP_\theta(x^n,z^n) \equiv \bar{D}_\theta(\bar{C}^{n,n}).
\label{eq:term1_1}
\end{equation}
By construction of $\cU$ and $C^{n,n}$, $(x^n,z^n) \in \cU$ implies that at least $\delta
n$ components of $\wh{\bar{x}}^n \equiv \bar{C}^{n,n}(x^n,z^n)$ satisfy
$\rho(x_i,\wh{\bar{x}}_i) > M$, so by definition of $\rho_M$ it follows that
$\rho_M(x^n,\bar{C}^{n,n}(x^n,z^n)) \ge n\delta M$ for all $(x^n,z^n) \in \cU$. Thus,
$$
\frac{1}{n}
\int_\cU \rho_M(x^n,\bar{C}^{n,n}(x^n,z^n))dP_\theta(x^n,z^n) \ge
\delta M \cdot P^n_\theta \times P^n_\theta (\cU),
$$
which, together with (\ref{eq:term1_1}), implies that
\begin{equation}
P^n_\theta \times P^n_\theta(\cU) \le \frac{\bar{D}_\theta(\bar{C}^{n,n})}{\delta M}.
\label{eq:term1_2}
\end{equation}
Using the Cauchy--Schwarz inequality, (\ref{eq:refletter}), and (\ref{eq:term1_2}), we can write
\begin{eqnarray}
\frac{1}{n}\int_\cU \rho(x^n,C^{n,n}(x^n,z^n))dP_\theta(x^n,z^n) &=&
\frac{1}{n}\E_\theta\left[\rho(X^n,a^n_*) \cdot 1_\cU \right]
\nonumber\\
&\le&
\sqrt{P^n_\theta \times P^n_\theta(\cU) \E_\theta[\rho^2(X^n,a^n_*)/n^2]} \nonumber\\
&\le& \sqrt{\frac{2G\bar{D}_\theta(\bar{C}^{n,n})}{\delta M}},
\label{eq:term1_3}
\end{eqnarray}
where the last inequality follows from the easily established fact that, for any $n$ independent random variables $V_1,\cdots,V_n$ satisfying $\E[V_i] \le G$, $\E\left[\left(\frac{1}{n}\sum^n_{i=1}V_i\right)^2\right] \le 2G$.

Now, $(x^n,z^n) \in \cU^c$ implies that for each $i=1,\cdots,n$ either
$\wh{x}_i = \wh{\bar{x}}_i$ and $\rho_M(x_i,\wh{x}_i) =
\rho(x_i,\wh{x}_i)$, or $\wh{x}_i = a_*$ and $\rho(x_i,\wh{\bar{x}}_i)
> M$, where $\wh{\bar{x}}^n \equiv
(\wh{\bar{x}}_1,\cdots,\wh{\bar{x}}_n) = \bar{C}^{n,n}(x^n,z^n)$ and
$\wh{x}^n \equiv (\wh{x}_1,\cdots,\wh{x}_n) = C^{n,n}(x^n,z^n)$. Then, by the union bound,
\begin{eqnarray}
&& \frac{1}{n}\int_{\cU^c} \rho(x^n,C^{n,n}(x^n,z^n))dP_\theta(x^n,z^n) =
\frac{1}{n} \sum^n_{i=1} \int_{\cU^c} \rho(x_i,\wh{x}_i)
dP_\theta(x^n,z^n) \nonumber\\
&& \qquad \le \frac{1}{n} \sum^n_{i=1}\int_{\cU^c} \rho_M(x_i,\wh{\bar{x}}_i)
dP_\theta(x^n,z^n) + \frac{1}{n} \sum^n_{i=1} \int_{\{(x^n,z^n) :
  \rho(x_i,\wh{\bar{x}}_i) > M \}} \rho(x_i,a_*) dP_\theta(x^n,z^n).
\label{eq:term2_1}
\end{eqnarray}
The first term on the right-hand side of (\ref{eq:term2_1}) is
bounded as
\begin{equation}
\frac{1}{n} \sum^n_{i=1}\int_{\cU^c} \rho_M(x_i,\wh{\bar{x}}_i)
dP_\theta(x^n,z^n) \le \bar{D}_\theta(\bar{C}^{n,n}).
\label{eq:term2_1a}
\end{equation}
As for the
second term, we can once again invoke the Cauchy--Schwarz inequality
and (\ref{eq:refletter}) to write
\begin{eqnarray}
\frac{1}{n}\sum^n_{i=1} \int_{\{(x^n,z^n) :
  \rho(x_i,\wh{\bar{x}}_i) > M \}} \rho(x_i,a_*)
dP_\theta(x^n,z^n) &\le& \frac{1}{n}\sum^n_{i=1} \sqrt{P_\theta(\{(x^n,z^n): \rho(x_i,\wh{\bar{x}}_i) > M \})
  \E_\theta[\rho^2(X,a_*)]} \nonumber\\
&\le& \frac{\sqrt{G}}{n}
\sum^n_{i=1}\sqrt {P_\theta(\{(x^n,z^n) :
  \rho(x_i,\wh{\bar{x}}_i) > M\})}.
\label{eq:term2_2}
\end{eqnarray}
Let us estimate the summation on the right-hand side of (\ref{eq:term2_2}). First of all,
note that
\begin{equation}
\frac{1}{n}\sum^n_{i=1}\int_{\{(x^n,z^n) : \rho(x_i,\wh{\bar{x}}_i) > M
  \}} \rho_M(x_i,\wh{\bar{x}}_i) dP_\theta(x^n,z^n) \le
\bar{D}_\theta(\bar{C}^{n,n}).
\label{eq:prob_1}
\end{equation}
Now, $\rho(x_i,\wh{\bar{x}}_i) > M$ implies that
$\rho_M(x_i,\wh{\bar{x}}_i) = M$, so that
$$
\int_{\{(x^n,z^n): \rho(x_i, \wh{\bar{x}}_i) > M \}}
\rho_M(x_i,\wh{\bar{x}}_i) dP_\theta(x^n,z^n) = M P_\theta(\{(x^n,z^n):
\rho(x_i,\wh{\bar{x}}_i) > M\}),
$$
which, together with (\ref{eq:prob_1}), yields the estimate
$$
\sum^n_{i=1} P_\theta(\{(x^n,z^n) : \rho(x_i,\wh{\bar{x}}_i) >
M \}) \le \frac{n \bar{D}_\theta(\bar{C}^{n,n})}{M},
$$
whence by the concavity of the square root it follows that
$$
\sum^n_{i=1} \sqrt{P_\theta(\{(x^n,z^n): \rho(x_i,\wh{\bar{x}}_i) > M
  \})} \le n \sqrt{\frac{\bar{D}_\theta(\bar{C}^{n,n})}{M}}.
$$
Substituting this bound into (\ref{eq:term2_2}) yields
\begin{equation}
\frac{1}{n} \sum^n_{i=1} \int_{\{(x^n,z^n): \rho(x_i,\wh{\bar{x}}_i) >
  M\}} \rho(x_i,a_*)dP_\theta(x^n,z^n) \le \sqrt{\frac{G \bar{D}_\theta(\bar{C}^{n,n})}{M}}.
\label{eq:term2_3}
\end{equation}
The lemma is proved by combining (\ref{eq:barmismatch}),
(\ref{eq:term1_3}), (\ref{eq:term2_1a}), and (\ref{eq:term2_3}).

\bibliography{ucm_ver2}

\end{document}